\documentclass[lettersize,journal]{IEEEtran}
\usepackage[T1]{fontenc}
\usepackage[utf8]{inputenc}
\ifdefined\DeclareUnicodeCharacter
\DeclareUnicodeCharacter{2191}{\ensuremath{\uparrow}}
\DeclareUnicodeCharacter{2193}{\ensuremath{\downarrow}}
\fi
\usepackage{needspace}
\usepackage{hhline}

\usepackage{amsmath,amsfonts,amssymb}
\usepackage{algorithmic}
\usepackage{algorithm}
\usepackage{bm}

\usepackage[table]{xcolor} 
\usepackage{graphicx}
\usepackage[caption=false,font=normalsize,labelfont=sf,textfont=sf]{subfig}
\usepackage{stfloats}
\usepackage{booktabs}
\usepackage{array}
\usepackage{multirow}
\usepackage{makecell}
\usepackage{stackengine}
 
\usepackage{textcomp}
\usepackage{url}
\usepackage{verbatim}
\usepackage{cite}

\newcommand\xrowht[2][0]{\addstackgap[.5\dimexpr#2\relax]{\vphantom{#1}}}
\newcommand{\hnxu}[1]{\textcolor{black}{#1}}

\newcommand{\revision}[1]{{\color{black}#1}}
\definecolor{syncpurple}{RGB}{128,0,128}
\newcommand{\syncfix}[1]{{\color{black}#1}}
\newcommand{\seedresult}[2]{\ensuremath{#1\pm #2}}
\newcommand{\dynamicpenalty}{DP}
\newcommand{\augmentedpenalty}{AL}

\begin{document}
\bstctlcite{IEEEexample:BSTcontrol}
\title{Towards High-performance and Fine-grained SSL Speech Foundation Models Pruning Using Sparsity-Aware Self-Pinching Gates}

\author{Haoning Xu, Zhaoqing Li, Youjun Chen, Huimeng Wang, Zengrui Jin, Guinan Li, Jiajun Deng, Chengxi Deng,  Tianzi Wang, Shujie Hu, Mingyu Cui, Mengzhe Geng, Xunying Liu,~\IEEEmembership{Member,~IEEE}
\thanks{Haoning Xu, Zhaoqing Li, Youjun Chen, Huimeng Wang, Guinan Li, Jiajun Deng, Chengxi Deng, Tianzi Wang, Shujie Hu, Mingyu Cui are with the Chinese University of Hong Kong, China (email: \{hnxu, zqli, yjchen, huimengwang, gnli, jjdeng, cxdeng, twang, sjhu, mycui\}@se.cuhk.edu.hk)}
\thanks{Zengrui Jin is with the Tsinghua University, China (email: zrjin@tsinghua.edu.cn)}
\thanks{Mengzhe Geng is with the National Research Council Canada, Canada (email: Mengzhe.Geng@nrc-cnrc.gc.ca);}
\thanks{Xunying Liu is with the Chinese University of Hong Kong, China and the corresponding author (email:  xyliu@se.cuhk.edu.hk).}
}

\markboth{IEEE Transactions on Audio, Speech and Language Processing}%
{Xu \MakeLowercase{\textit{et al.}}: Sparsity-Aware Self-Pinching Gates}
\AddToHookNext{shipout/foreground}{%
  \put(49bp,-758bp){\makebox(0,0)[lt]{%
    \parbox{514bp}{\fontsize{7}{8}\selectfont
    \textcopyright\ 2026 IEEE. Personal use of this material is permitted. Permission from IEEE must be obtained for all other uses, in any current or future media, including reprinting/republishing this material for advertising or promotional purposes, creating new collective works, for resale or redistribution to servers or lists, or reuse of any copyrighted component of this work in other works.}%
  }}%
}

\maketitle

\begin{abstract}
This paper presents a high-performance and scalable approach for speech foundation model compression that tightly integrates model pruning and parameter update into one consistent cycle. Highly compact sparsity-aware self-pinching gates (SASPG) each containing a single learnable threshold are layer-level tied. They are jointly learned with uncompressed models using either CTC cost for ASR task-specific model pruning, or distillation loss in task-agnostic pruning. Both fine-grained node-level unstructured, and coarse-grained channel-level structured pruning are designed to flexibly adjust the trade-off between performance and hardware-friendly speedup. 
Extensive experiments conducted on the LibriSpeech-100hr, LibriSpeech-960hr and SUPERB datasets suggest consistent performance improvements over parameter magnitude-based offline pruning, \hnxu{coarse-grained} NAS and DPHuBERT-based integrated pruning approaches.
Average ASR word error rate (WER) reductions of up to 51.98\% and 7.61\% absolute (76.32\% and 46.04\% relative) are obtained using SASPG across the LibriSpeech \hnxu{dataset} ({\tt dev-clean}, {\tt dev-other}, {\tt test-clean}, and {\tt test-other}) over parameter magnitude pruned wav2vec2.0-Base and HuBERT-Large models, respectively. 
\hnxu{Node-level SASPG achieves statistically lossless 5x (80\% sparsity) and 3.3x (70\% sparsity) \syncfix{compression ratios} on wav2vec2.0-Base and HuBERT-Large, respectively.} \revision{Channel-level SASPG achieves maximum average WER reductions of 13.50\% and 2.47\% absolute over its magnitude-based counterparts on wav2vec2.0-Base and HuBERT-Large, respectively. Measured speedups reach up to 1.62$\times$ and 1.87$\times$, respectively.}
SASPG's performance superiority over \hnxu{magnitude-based pruning} and DPHuBERT is widely demonstrated across \hnxu{9, 8, 9 and 8} SUPERB subtasks out of 10 on wav2vec2.0-Base, HuBERT-Base, WavLM-Base+ and HuBERT-Large models, respectively, \hnxu{while achieving a substantial 2056x reduction in pruning-specific parameter overhead compared to DPHuBERT. Our code is available at \url{https://github.com/hanyxu/SASPG-pruning}.}
\end{abstract}

\begin{IEEEkeywords}
Model pruning, speech foundation models, speech recognition, knowledge distillation.
\end{IEEEkeywords}

\section{Introduction}

\IEEEPARstart{T}{he} {development of speech processing has shifted from fully supervised architectures \cite{bahdanau2016end, gulati2020conformer, chan2016listen,chorowski2015attention,hori2017joint, amodei2016deep, prabhavalkar2023end} to self-supervised learning (SSL) frameworks~\cite{baevski2022data2vec,baevskivq2020vqwav2vec,baevski2020wav2vec,hsu2021hubert,chen2022wavlm}. 
Representative SSL models, including wav2vec 2.0~\cite{baevski2020wav2vec}, HuBERT~\cite{hsu2021hubert}, and WavLM~\cite{chen2022wavlm}, learn transferable speech representations from large-scale unlabeled audio and achieve superior performance when adapted to downstream tasks using limited labeled data.}

These advances, however, incur substantial memory and computational costs that limit real-time inference for downstream tasks (e.g., ASR) and on-device deployment, motivating the effective compression of SSL speech foundation models. 
Existing compression techniques can be broadly grouped into low-bit quantization, which replaces floating-point weights with low-precision values~\cite{uq-2bcfm,uq-4bcfm,ibert,xu2025effective,mq-person, usmlite, li2025towards, egiazarian2024extreme, gao2021extremely, onepass-zq, liu2025mixpq, mp-w2v, mq-jh, mq-jh-1, peng2021shrinking, wagner2024outlier, xu2022towards-jh, zhao2024ada, zhen2023sub8cfm}; and architecture compression~\cite{wang2022lighthubert,fithubert,li25v_interspeech,li2023lossless, wang2024fact, lr1, distilhubert, chang2023colld, cho2023sd, distw2v, Moequant, park2023conformer, shao2024dq, zampierin2024skillsimilarityawareknowledgedistillation, gu2024sparsewav,xu25e_interspeech, peng2023dphubert,wang2023task,hj
,pru5,fu2022losses,lodagala2023pada, lai2021parp, yang2023learning}, which reduces structural redundancy.

Architecture compression approaches can be broadly categorized into: 
\textbf{1) low-rank factorization}~\cite{li2023lossless, wang2024fact, lr1}, which factorizes a large weight matrix into two smaller factors, with one often constrained to be semi-orthogonal ~\cite{li2023lossless}. 
\textbf{2) knowledge distillation (KD)}~\cite{distilhubert, peng2023dphubert, chang2023colld, cho2023sd, distw2v, Moequant, park2023conformer, shao2024dq,zampierin2024skillsimilarityawareknowledgedistillation}, which compresses a large pretrained teacher model into a compact student model by 
aligning the output or intermediate representation between teacher and student models.
\textbf{3) model pruning}~\cite{gu2024sparsewav,xu25e_interspeech, peng2023dphubert,wang2023task,hj, pru5, fu2022losses, lodagala2023pada, lai2021parp, yang2023learning}, which can be further divided into: 
\textbf{node-level unstructured pruning}~\cite{gu2024sparsewav,xu25e_interspeech,fu2022losses,siyuan}, which removes individual weights and can achieve high compression ratios, although the resulting irregular sparse matrices often require specialized libraries for inference acceleration; and
\textbf{channel-level structured pruning}~\cite{peng2023dphubert,wang2023task,hj, siyuan}, which removes entire components, such as attention heads, and therefore provides direct hardware-friendly acceleration. 
However, pruning aggregated structural groups can remove functionally important parameters and cause substantial performance degradation.

Focusing specifically on model pruning, recent studies have mainly explored two directions.
The first~\cite{gu2024sparsewav,liu2025mixpq, hj, jiang2023accurate, wang2023task} develops weight-importance criteria \cite{L0reg, OBC, OBS, maddison2022concrete} to identify and retain critical parameters while removing redundant ones. 
The second combines pruning with other compression techniques, such as KD~\cite{peng2023dphubert}, quantization~\cite{usmlite,onepass-zq} or low-rank factorization \cite{wang2024fact,li2023lossless} to improve compression while preserving performance.
\syncfix{Joint optimization of pruning thresholds and model weights has also been explored using differentiable mask approximations~\cite{azarian2020learned_threshold_pruning,liu2020dynamic_sparse_training}. For SSL speech models, learnable gates with sparsity constraints have been applied to task-specific compression for speaker verification, anti-spoofing, and diarization~\cite{peng2026hybrid_pruning,han2026robust_diarization_pruning}.}

Despite this progress, pruning SSL-based models remains challenging:
\textbf{1)} \textbf{Pruning can cause severe performance degradation} when salient parameters are mistaken for redundant ones, or when the removal of coarse-grained structural units inadvertently discards essential functional components \cite{wang2023task,peng2023dphubert,hj,zampierin2024skillsimilarityawareknowledgedistillation}. Additionally, few studies specify statistically grounded criteria for distinguishing acceptable from unacceptable performance loss \cite{gillick1989some}.
\textbf{2) Pruning and parameter update are often performed in separate stages}, which decouples pruning target selection from parameter optimization \cite{lai2021parp, yang2023learning, lodagala2023pada, gu2024sparsewav}. This disjointed paradigm forces the compression process into a sub-optimal trajectory where pruning decisions are not dynamically guided by the target task objective.
\textbf{3) Difficulty in achieving a flexible trade-off between coarse- and fine-grained pruning.} 
Methods that assign learnable variables to structural units usually impose a one-to-one mapping between units and auxiliary parameters \cite{hj, peng2023dphubert, wang2023task}. 
Extending this design to node-level pruning introduces pruning-specific parameters that scale with the total number of weights~\cite{fu2022losses}.
\textbf{4) Many pruning pipelines are designed for a single downstream task}, such as ASR~\cite{gu2024sparsewav, xu25e_interspeech, usmlite}. 
Adapting the compressed model to another downstream application then requires a new pruning and fine-tuning process~\cite{hj,wang2024fact}, which limits reuse across tasks.

To this end, this paper proposes \textbf{sparsity-aware self-pinching gates (SASPG)}, a compact and scalable framework for compressing SSL speech foundation models. 
SASPG integrates pruning and parameter update into a single optimization process. 
\syncfix{The method uses learnable thresholds shared within each pruned module to construct pruning masks from current weight magnitudes or structural-unit scores, keeping the pruning-specific parameter overhead low.} 
It can be trained with a CTC loss for ASR-specific pruning or with a distillation loss for task-agnostic pruning. 
The framework supports unstructured pruning for performance-oriented compression and structured pruning for hardware-friendly acceleration.

Experiments conducted on LibriSpeech \cite{panayotov2015librispeech} and SUPERB \cite{superb} present consistent improvements over magnitude-based offline pruning, coarse-grained channel-level NAS \cite{xu25e_interspeech}, and DPHuBERT-based joint pruning \cite{peng2023dphubert}. 
\textbf{1)} On LibriSpeech dev-clean, dev-other, test-clean, and test-other, SASPG achieves average WER reductions of up to \syncfix{51.98 and 7.61 absolute percentage points} (76.32\% and 46.04\% relative) over magnitude-pruned wav2vec2.0-Base and HuBERT-Large models, respectively. 
SASPG also outperforms the NAS baseline across all evaluated sparsity levels. 
\textbf{2)} Node-level SASPG achieves maximum lossless \syncfix{compression ratios} of 5x (80\% sparsity) and 3.3x (70\% sparsity) on wav2vec2.0-Base and HuBERT-Large, respectively, with no statistically significant WER increase. 
\textbf{3)} Channel-level \syncfix{SASPG} \revision{achieves measured speedups of up to 1.62x on wav2vec2.0-Base and 1.87x on HuBERT-Large}, \revision{with maximum average WER reductions of 13.50\% and 2.47\% absolute compared with magnitude-based structured pruning.} 
\textbf{4)} Task-agnostic pruning via distillation shows that \syncfix{SASPG across its node- and channel-level configurations outperforms} magnitude-based pruning and DPHuBERT on 9, 8, 9, and 8 of 10 SUPERB subtasks for wav2vec2.0-Base, HuBERT-Base, WavLM-Base+, and HuBERT-Large, respectively, while reducing pruning-specific parameter overhead by up to 2056x compared with DPHuBERT.

The main contributions of this paper are as follows:

\begin{itemize}
    \item \textbf{Consistency between pruning and parameter update.} SASPG replaces the sequential prune-then-update workflow of uniform magnitude-based pruning (UMP) with a joint optimization process. 
    By comparing current weight magnitudes or structural scores with learnable thresholds, SASPG produces differentiable probabilities and enables layer-wise sparsity allocation during training.
    
    \item \textbf{Flexible pruning granularity.} 
    Existing channel-level methods rely on coarse structural units and introduce auxiliary parameters tied to candidate masks or structural components. 
    SASPG employs a unified threshold-based formulation that readily generalizes from node-level unstructured to channel-level structured pruning at any desired granularity.
    
    \item \textbf{Task-agnostic pruning via distillation.} 
    Beyond our previous ASR-specific pruning~\cite{xu25e_interspeech}, SASPG is extended to task-agnostic pruning by replacing the CTC objective with a hidden-state distillation loss. Crucially,  integrating differentiable target sparsity into the distillation objective eliminates the empirical penalty scaling in~\cite{xu25e_interspeech}. Experiments on the SUPERB benchmark demonstrate its effectiveness across diverse downstream tasks.
\end{itemize}

The rest of the paper is organized as follows. Section~\ref{sec2} provides an overview of the key speech foundation models relevant to this work. Section~\ref{sec3} introduces the preliminaries, including {notational} definitions and pruning granularity. Section~\ref{ASR section} details the proposed SASPG method alongside baseline methods for ASR-specific foundation {models} pruning. Section~\ref{sec5} extends the pruning frameworks to task-agnostic scenarios utilizing knowledge distillation. Section~\ref{sec6} presents extensive experimental setups, results, and analyses across both the ASR-specific and SUPERB benchmarks. Finally, Section~\ref{sec7} concludes the paper.

\section{Speech Foundation Models}
\label{sec2}

\begin{figure}[htbp]
    \centering
    \includegraphics[scale=0.55]{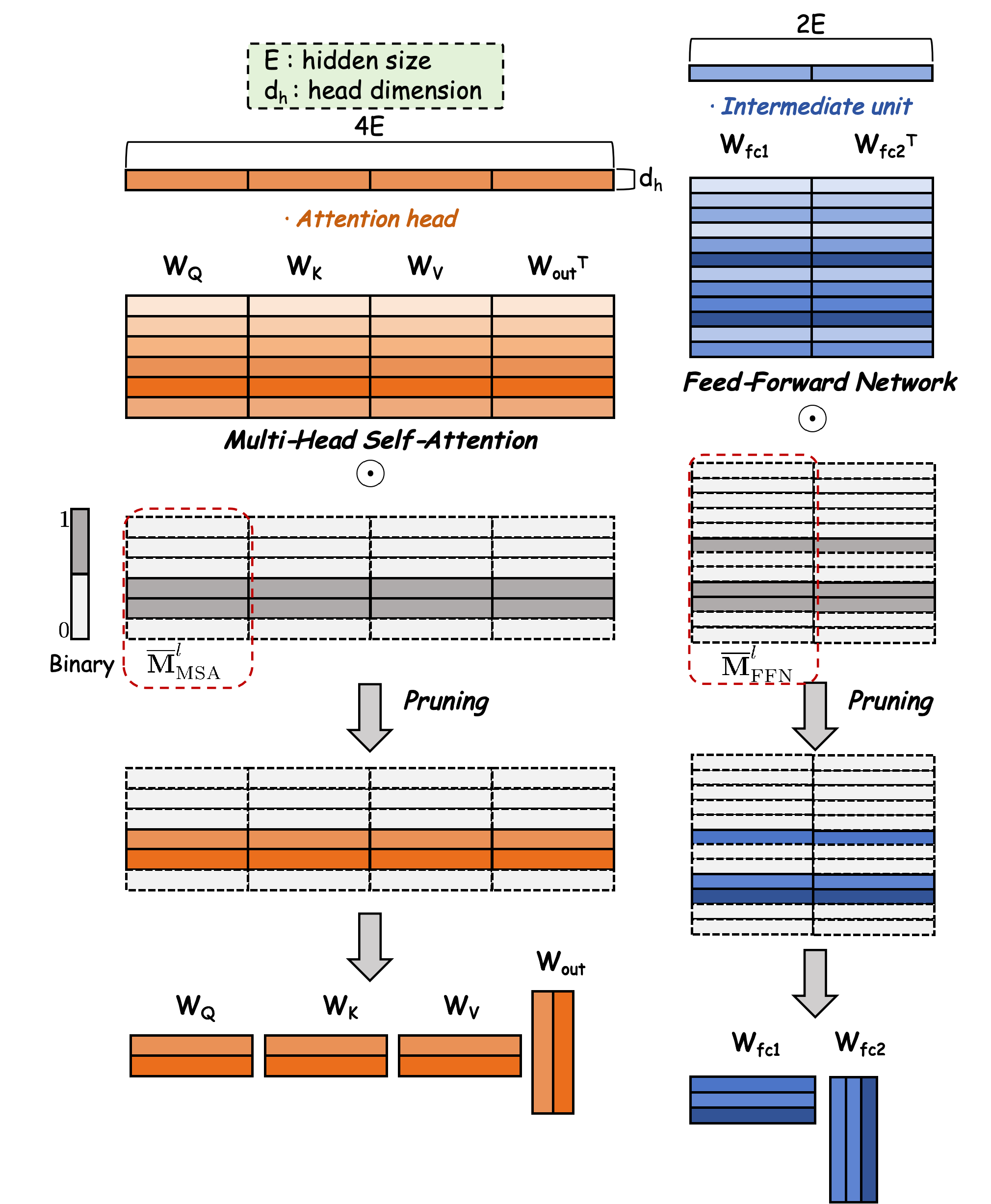}
    \caption{Minimal structural units considered for channel-level structured pruning: attention heads (left) and feed-forward intermediate neurons (right).}
    \label{Fig:units}
\end{figure}

The field of speech processing has experienced a fundamental transition from traditional fully supervised architectures to powerful Self-Supervised Learning (SSL) paradigms. Initially, high-performance recognition relied on fully supervised models such as the Conformer~\cite{gulati2020conformer}, which intelligently fuses convolutional neural networks with self-attention to capture both local and global acoustic patterns. However, training these models from scratch requires thousands of hours of high-quality transcribed audio, creating a formidable data bottleneck. By leveraging large amounts of unlabeled audio, SSL effectively resolves this limitation.

Foundational models such as wav2vec 2.0~\cite{baevski2020wav2vec} pioneered this approach using contrastive learning on the LibriSpeech corpus~\cite{panayotov2015librispeech}. Subsequently, HuBERT~\cite{hsu2021hubert} advanced the paradigm by adopting a masked prediction objective similar to BERT~\cite{devlin-etal-2019-bert} in natural language processing. Scaled to the larger Libri-light dataset~\cite{kahn2020libri}, HuBERT learns to predict discrete pseudo-labels generated through offline acoustic clustering. Building upon these successes, WavLM~\cite{chen2022wavlm} further extends this framework across diverse data \cite{chen21o_interspeech, wang2021voxpopuli}. By incorporating speech denoising objectives, WavLM establishes state-of-the-art robustness across various acoustic environments, as demonstrated on universal evaluations such as SUPERB ~\cite{superb}.

However, a universal trend in this SSL evolution is the continuous scaling of model capacity to achieve superior performance. These foundation models typically feature sizable parameter counts, adopting Base ($\sim$95M parameters) and Large ($\sim$317M parameters) configurations. While \hnxu{these models ensure} strong generalization, this immense footprint creates critical memory and computational bottlenecks. Their scale renders direct deployment on resource-constrained edge devices computationally infeasible. Consequently, effective architecture compression for speech SSL models has emerged as an urgent necessity for real-world deployment.

\section{Preliminaries}
\label{sec3}

Despite differences in training objectives, these SSL models share a unified backbone architecture comprising a Convolutional Neural Network (CNN) feature extractor and a Transformer encoder. The Transformer encoder is the dominant component, accounting for over 90\% of the total parameters. For instance, in the HuBERT-Large model, the encoder consists of 24 blocks, where each block contains a \textbf{M}ulti-Head \textbf{S}elf-\textbf{A}ttention (MSA) module and a \textbf{F}eed-\textbf{F}orward \textbf{N}etwork (FFN) module.

As shown in Fig.~\ref{Fig:units}, this pruning process occurs exclusively within the SSL encoder, specifically targeting the linear layers within the MSA (matrices $\{\mathbf{W}_q, \mathbf{W}_k, \mathbf{W}_v, \mathbf{W}_{out}\}$) and the FFN (matrices $\{\mathbf{W}_{fc1}, \mathbf{W}_{fc2}\}$).
Other model components, including the CNN feature extractor, remain unpruned.

\subsection{Notation Definition}
Consider a Transformer encoder of an SSL model with $L$ layers, where each layer contains a MSA and a FFN module. To simplify the notation, we introduce a placeholder subscript $*$ to denote any specific type of linear projection matrix. We define the projection sets for the two modules as $\mathcal{P}_{\text{MSA}} = \{q, k, v, out\}$ and $\mathcal{P}_{\text{FFN}} = \{fc1, fc2\}$. The universal projection set is thus explicitly $\mathcal{P} = \mathcal{P}_{\text{MSA}} \cup \mathcal{P}_{\text{FFN}} = \{q, k, v, out, fc1, fc2\}$.

For the $l$-th transformer layer ($1 \le l \le L$), we define the original weight matrix set $\mathcal{W}^l$, along with its corresponding soft mask matrix set $\widetilde{\mathcal{M}}^l$ and hard (binary) mask matrix set $\overline{\mathcal{M}}^l$, compactly as:
\begin{equation}
\mathcal{W}^l = \{\mathbf{W}_{*}^{l}\}, \quad
\widetilde{\mathcal{M}}^l = \{\widetilde{\mathbf{M}}_{*}^{l}\}, \quad
\overline{\mathcal{M}}^l = \{\overline{\mathbf{M}}_{*}^{l}\},
\quad {*\in\mathcal{P}}.
\end{equation}

The properties of the individual mask matrices are defined as follows:
\begin{itemize}
    \item \textbf{Soft mask} $\widetilde{\mathbf{M}}_{*}^{l} \in \widetilde{\mathcal{M}}^l$: Elements are continuous floating-point numbers in $[0,1]$.
    \item \textbf{Hard mask} $\overline{\mathbf{M}}_{*}^{l} \in \overline{\mathcal{M}}^l$: Elements are discrete binary values $\{0, 1\}$. 
\end{itemize}

Correspondingly, for each specific projection matrix type $* \in \mathcal{P}$, the soft and hard masked weight matrices are strictly constructed via the Hadamard product (element-wise multiplication $\odot$):
\begin{equation}
\widetilde{\mathbf{W}}_{*}^{l} = \widetilde{\mathbf{M}}_{*}^{l} \odot \mathbf{W}_{*}^{l}, \quad
\overline{\mathbf{W}}_{*}^{l} = \overline{\mathbf{M}}_{*}^{l} \odot \mathbf{W}_{*}^{l}.
\end{equation}
Their corresponding sets for the $l$-th transformer layer are denoted as $\widetilde{\mathcal{W}}^l = \{\widetilde{\mathbf{W}}_{*}^{l}\}_{*\in\mathcal{P}}$ and $\overline{\mathcal{W}}^l = \{\overline{\mathbf{W}}_{*}^{l}\}_{*\in\mathcal{P}}$.

Finally, global sets for the entire network are denoted by simply omitting the layer superscript $l$ (e.g., $\mathcal{W} = \bigcup_{l=1}^L \mathcal{W}^l$, and similarly for $\overline{\mathcal{M}}, \widetilde{\mathcal{M}}, \overline{\mathcal{W}}, \widetilde{\mathcal{W}}$).

\subsection{Sparsity Metrics} \label{sec:sparsity}
We define the retained parameter count (structural penalty) and sparsity ratio at both layer and global levels. For the $l$-th layer with \syncfix{$N^l$ parameters in the prunable matrices}, the retained parameter count is the $L_0$ norm of its binary masks, $\mathcal{R}(\overline{\mathcal{M}}^l) = \sum_{* \in \mathcal{P}} \|\overline{\mathbf{M}}_{*}^{l}\|_0$. The \textbf{layer-wise sparsity} is then $R_c(\overline{\mathcal{M}}^l) = 1 - \mathcal{R}(\overline{\mathcal{M}}^l)/N^l$.

Aggregating across all $L$ transformer layers yields the global retained parameter count $\mathcal{R}(\overline{\mathcal{M}}) = \sum_{l=1}^L \mathcal{R}(\overline{\mathcal{M}}^l)$. The overall \textbf{global sparsity} $R_c$ is defined as:
\begin{equation}
R_c = 1 - \frac{\mathcal{R}(\overline{\mathcal{M}})}{N_{\text{total}}},
\end{equation}
where \syncfix{$N_{\text{total}}=\sum_{l=1}^{L}N^l$ counts the parameters in the same prunable matrices before pruning}.

\subsection{Pruning Granularity} \label{granularity}
Based on the defined notation, we adopt two pruning strategies:
\begin{itemize}
    \item \textbf{Node-level Unstructured Pruning:} Under this paradigm, the elements within any hard mask $\overline{\mathbf{M}}_{*}^{l} \in \overline{\mathcal{M}}^l$ are independently determined. The pruning operation is performed at the fine-grained level of individual scalar parameters, resulting in sparse matrices with irregular patterns. All bias terms remain unpruned.
    
    \item \textbf{Channel-level Structured Pruning:} To ensure hardware acceleration and structural consistency, we define a \textbf{structural unit} as a \textbf{coupled group of channels} spanning multiple matrices, serving as the minimal pruning unit for all modules $m \in \{\text{MSA}, \text{FFN}\}$. To formally encapsulate this, we introduce a hard \textbf{module-level mask} $\overline{\mathbf{M}}_{m}^{l}$ to dictate the shared sparsity pattern. As illustrated in Fig.~\ref{Fig:units}, pruning the $j$-th unit in $\overline{\mathbf{M}}_{m}^{l}$ simultaneously zeroes out its constituent channels across the coupled individual masks $\overline{\mathbf{M}}_{*}^{l}$ ($* \in \mathcal{P}_m$). The constraints are defined as follows:
    
    \textbf{1) Attention Head ($m = \text{MSA}$):} The $j$-th head is a unit comprising the $j$-th block of $d$ channels ($d$ being the head dimension). If pruned, the corresponding mask blocks are strictly coupled:
    \begin{equation}
    \begin{aligned}
    \overline{\mathbf{M}}_q^l\big|_{\syncfix{(j-1)\cdot d : j\cdot d},\ :} &= \overline{\mathbf{M}}_k^l\big|_{\syncfix{(j-1)\cdot d : j\cdot d},\ :} \\
    &= \overline{\mathbf{M}}_v^l\big|_{\syncfix{(j-1)\cdot d : j\cdot d},\ :} \\
    &= (\overline{\mathbf{M}}_{out}^l)^T\big|_{\syncfix{(j-1)\cdot d : j\cdot d},\ :} = \mathbf{0}.
    \end{aligned}
    \end{equation}
    The input-side biases ($q$, $k$, $v$ projections) are pruned alongside their respective weight blocks.
    
    \textbf{2) FFN Intermediate Neuron ($m = \text{FFN}$):} The $j$-th neuron is a unit comprising the $j$-th channel. If pruned, the corresponding mask rows are strictly coupled:
    \begin{equation}
    \overline{\mathbf{M}}_{fc1}^l\big|_{j,\ :} = (\overline{\mathbf{M}}_{fc2}^l)^T\big|_{j,\ :} = \mathbf{0}.
    \end{equation}
    The input-side bias ($fc1$ projection) is pruned alongside its weight row.
    Consequently, for any specific matrix $\mathbf{W}_{*}^{l}$ in module $m$, the effective structurally pruned weights are directly derived via element-wise multiplication:
    \begin{equation}
    \overline{\mathbf{W}}_{*}^{l} = 
    \begin{cases} 
    (\overline{\mathbf{M}}_{m}^{l})^T \odot \mathbf{W}_{*}^{l}, & \text{if } * \in \{out, fc2\}, \\
    \overline{\mathbf{M}}_{m}^{l} \odot \mathbf{W}_{*}^{l}, & \text{otherwise}.
    \end{cases}
    \end{equation}
    To be absolutely explicit, this formulation guarantees that the individual mask matrices within a given module strictly share the identical mask pattern (accounting for necessary transpositions): $\overline{\mathbf{M}}_{\text{MSA}}^{l} \equiv \overline{\mathbf{M}}_{q}^{l} = \overline{\mathbf{M}}_{k}^{l} = \overline{\mathbf{M}}_{v}^{l} = (\overline{\mathbf{M}}_{out}^{l})^T$, and similarly, $\overline{\mathbf{M}}_{\text{FFN}}^{l} \equiv \overline{\mathbf{M}}_{fc1}^{l} = (\overline{\mathbf{M}}_{fc2}^{l})^T$.
\end{itemize}

\section{ASR task-specific Foundation Model Pruning} 
\label{ASR section}

\subsection{Uniform Magnitude-based Pruning} \label{ump}
\begin{figure}[htbp]
    \centering
    \includegraphics[scale=0.25]{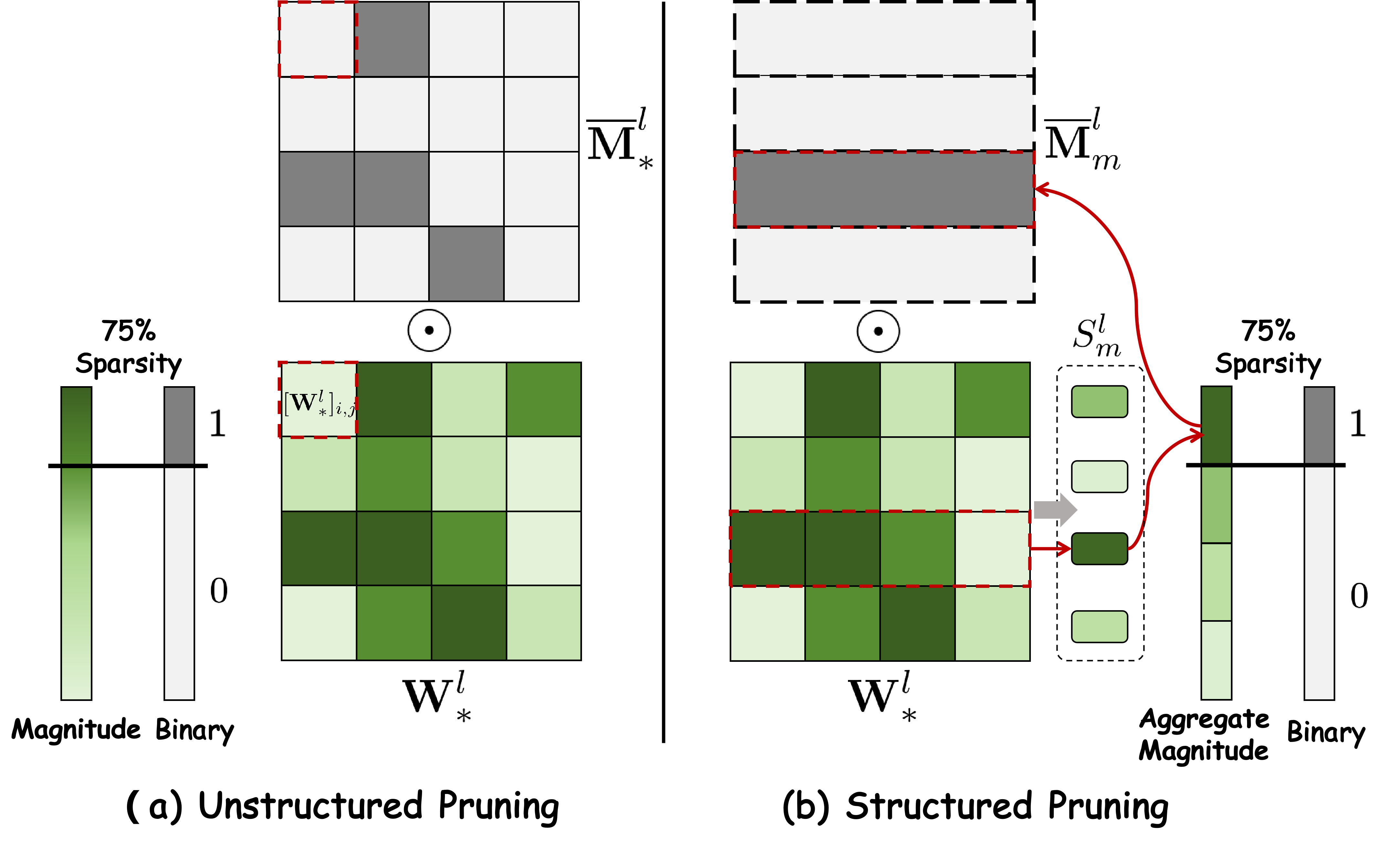}
    \caption{\textbf{Illustration of Uniform Magnitude-based Pruning (UMP).} (a) \textbf{Unstructured Pruning}: Individual scalar weights are evaluated independently by their absolute magnitudes $|[\mathbf{W}_{*}^{l}]_{i, j}|$, yielding irregular binary masks $\overline{\mathbf{M}}_{*}^{l}$. (b) \textbf{Structured Pruning}: Entire structural units are evaluated via aggregated magnitude scores $S_{m, j}^{l}$ (Eqs.~\ref{score_att}-\ref{score_ffn}). The lowest-scoring units are removed, producing the hard module-level mask $\overline{\mathbf{M}}_{m}^{l}$.}
    \label{fig:UMP}
\end{figure}

As illustrated in Fig.~\ref{fig:UMP}, Uniform Magnitude-based Pruning (UMP)~\cite{IMP} assumes that parameters or structural units with smaller magnitudes contribute less to the model's output. We apply this one-shot method, followed by fine-tuning for performance recovery, across two granularities:

\textbf{Unstructured Pruning.} The importance of each scalar parameter $[\mathbf{W}_{*}^{l}]_{i,j}$ in a weight matrix $\mathbf{W}_{*}^{l}$ ($* \in \mathcal{P}$) is defined by its absolute magnitude $|[\mathbf{W}_{*}^{l}]_{i,j}|$. Parameters are sorted layer-wise, and those below a predefined threshold are zeroed out, yielding sparse weight matrices with irregular patterns.

\textbf{Structured Pruning.} For attention heads and FFN intermediate neurons (Section~\ref{granularity}), unit importance is the \textbf{aggregate magnitude} of its constituent coupled weights. For the $j$-th \textbf{attention head} ($m=\text{MSA}$) in the $l$-th layer, the score $S_{\text{MSA}, j}^{l}$ sums the squared \syncfix{Frobenius norms} of the corresponding channel slices across aligned projection matrices:
\begin{equation}
S_{\text{MSA}, j}^{l} = \sum_{\mathbf{V} \in \{\mathbf{W}_q^l, \mathbf{W}_k^l, \mathbf{W}_v^l, (\mathbf{W}_{out}^l)^T\}} \lVert \mathbf{V}\big|_{\syncfix{(j-1)\cdot d : j\cdot d},\ :} \rVert_{\syncfix{F}}^2,
\label{score_att}
\end{equation}
where $d$ denotes the number of channels per attention head \syncfix{and $j=1,\ldots,N_{\mathrm{MSA}}$ indexes the heads using half-open channel slices}. Similarly, for the $j$-th \textbf{intermediate neuron} ($m=\text{FFN}$), the score $S_{\text{FFN}, j}^{l}$ aggregates the squared $L_2$-norms of its corresponding channels:
\begin{equation}
S_{\text{FFN}, j}^{l} = \sum_{\mathbf{V} \in \{\mathbf{W}_{fc1}^l, (\mathbf{W}_{fc2}^l)^T\}} \lVert \mathbf{V}\big|_{j,\ :} \rVert_2^2.
\label{score_ffn}
\end{equation}
Units with the lowest structural scores $S_{m, j}^{l}$ are then pruned based on a predetermined ratio.

During subsequent fine-tuning, the hard mask $\overline{\mathcal{M}}$ remains frozen. Thus, the objective function relies solely on the ASR-specific CTC loss $\mathcal{L}_{\text{ctc}}$, excluding any sparsity penalty.

However, a key limitation of UMP is that it enforces \textit{isotropic sparsity}---pruning an identical proportion of parameters across all layers. This ignores layer-specific sensitivities, as some layers can tolerate aggressive pruning while others degrade significantly even under conservative sparsity.

\subsection{NAS-based Structured Channel-wise Pruning}
\begin{figure}[htbp]
    \centering
    \includegraphics[scale=0.3]{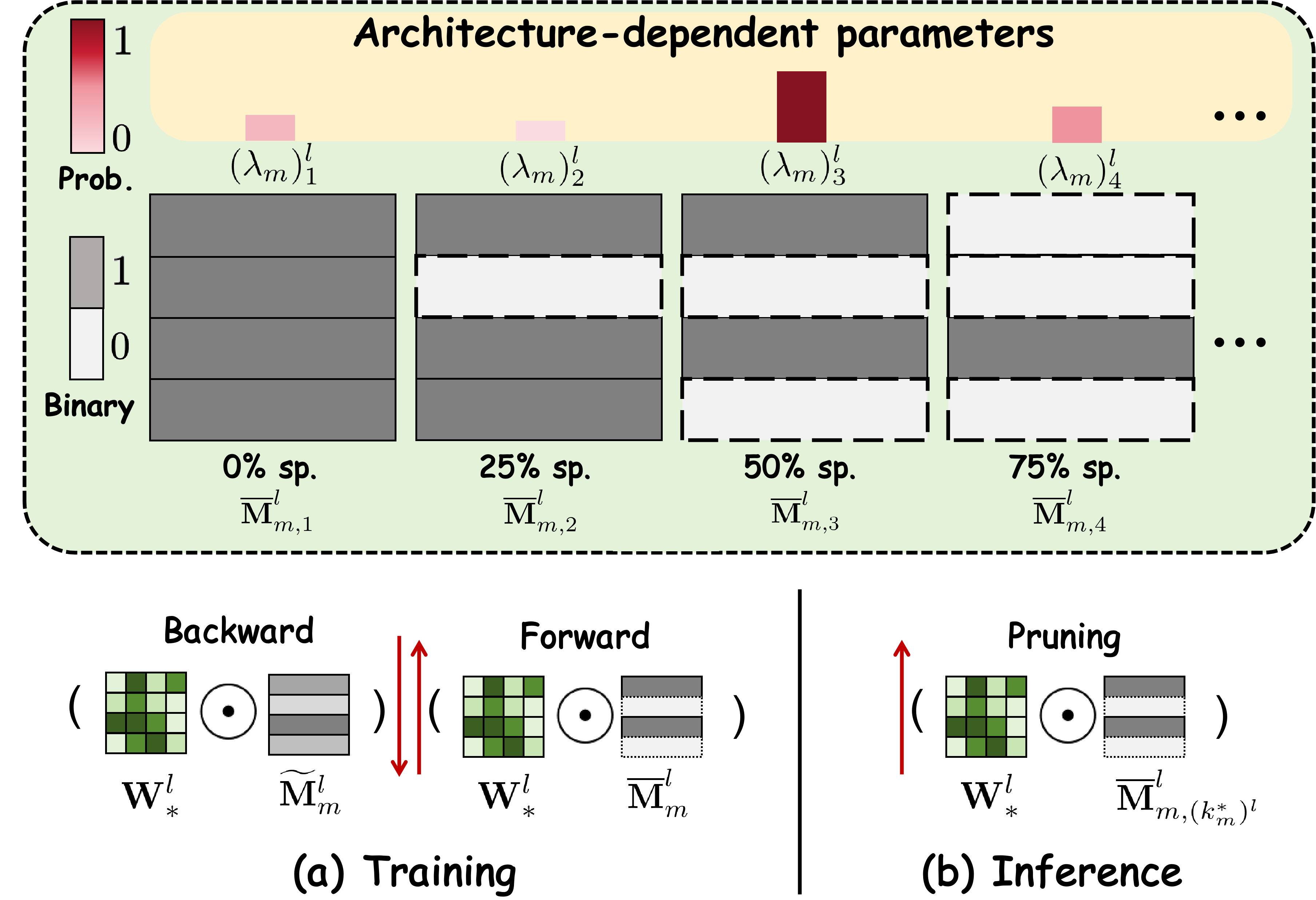}
    \caption{\textbf{Illustration of NAS-based Structured Channel-wise Pruning (NASP).} Candidate binary masks $\{\overline{\mathbf{M}}_{m, k}^l\}_{k=1}^7$ are \hnxu{pre-generated across} predefined sparsity ratios. During search, learnable architecture logits $(\alpha_m)_k^l$ yield soft categorical vectors $\bm{\lambda}_m^l$ for continuous backward gradient routing, while hard one-hot vectors $\overline{\bm{\lambda}}_m^l$ are sampled for discrete forward masking. At inference, the optimal candidate $(k_m^*)^l = \operatorname*{argmax}_k (\alpha_m)_k^l$ is deterministically selected to mask the dense weights.}
    \label{Fig:NASP}
\end{figure}

As shown in Fig.~\ref{Fig:NASP}, to enable the model to automatically learn the optimal structure, NAS-based Structured Channel-wise Pruning (NASP)~\cite{xu2025effective} employs a joint pruning and fine-tuning strategy. 

Building upon our unified notation and the global structural penalty $\mathcal{R}(\overline{\mathcal{M}})$ defined in Section~\ref{sec:sparsity}, the theoretical training objective minimizes both the ASR-specific CTC loss and the expected structural penalty:
\begin{equation}
\min_{\mathcal{W}, \bm{\alpha}} \mathbb{E}_{\overline{\mathcal{M}} \sim q_{\bm{\alpha}}} \left[ \frac{1}{D} \sum_{i=1}^D \mathcal{L}_{\text{ctc}}\left(f(x_i; \mathcal{W} \odot \overline{\mathcal{M}}), y_i\right) + \eta \cdot \mathcal{R}(\overline{\mathcal{M}}) \right],
\end{equation}

\textbf{\syncfix{Dynamic $\eta$-based Penalty (DP)}}. We dynamically adjust $\eta$ based on the deviation between the current global sparsity $R_c$ and the target sparsity $R_t$, using a tolerance margin of $\pm 0.5\%$:
\begin{equation}
\eta = 
\begin{cases} 
-\eta_0 & \text{if } R_{c} > R_{t} + 0.5\%, \\
0 & \text{if } R_{t} - 0.5\% \leq R_{c} \leq R_{t} + 0.5\%, \\
\eta_0 & \text{if } R_{c} < R_{t} - 0.5\%. \\
\end{cases} \label{eta}
\end{equation}

\textbf{Candidate Generation.} 
For each structural module $m \in \{\text{MSA}, \text{FFN}\}$ in the $l$-th transformer layer, we pre-generate $K=7$ candidate hard module-level masks $\{\overline{\mathbf{M}}_{m, k}^l\}_{k=1}^7$ by pruning the structural units with the lowest aggregated scores $S_{m, j}^{l}$, corresponding to a predefined set of sparsity ratios: $\{0\%, 25\%, 50\%, 75\%, 87.5\%, 90\%, 92.5\%\}$. 

\textbf{Training via Straight-Through Gumbel-Softmax.} 
To enforce physical sparsity, we use the Straight-Through (ST) Gumbel-Softmax trick. Let $(\pi_m)_k^l$ be the soft probability for the $k$-th candidate:
\begin{equation}
(\pi_m)_k^l = \frac{\exp(((\alpha_m)_k^l + G_k)/T)}{\sum_{j=1}^7 \exp(((\alpha_m)_j^l + G_j)/T)},
\end{equation}
where $(\alpha_m)_k^l$ are the learnable architecture logits, $T$ is the temperature parameter, and $G_k \sim \text{Gumbel}(0,1)$.

During the forward pass, a hard one-hot vector $\overline{\bm{\lambda}}_m^l = \text{one\_hot}(\arg\max_j (\pi_m)_j^l)$ (e.g., $[0,0,1,0,0,0,0]$) is sampled to select the hard module-level mask $\overline{\mathbf{M}}_{m}^{l} = \sum_{k=1}^7 (\overline{\lambda}_m)_k^l \cdot \overline{\mathbf{M}}_{m, k}^l$, which intrinsically yields the structurally pruned weights for module $m$, following the exact mechanism established in Section~\ref{granularity}.

During the backward pass, gradients are routed through the soft categorical vectors $\bm{\lambda}_m^l$ (where $\lambda_{m,k}^l = (\pi_m)_k^l$, e.g., $[0.1, 0.05, 0.7, 0.05, 0.05, 0.02, 0.03]$). Then the soft module-level mask is defined as $\widetilde{\mathbf{M}}_{m}^{l} = \sum_{k=1}^7 \lambda_{m,k}^l \cdot \overline{\mathbf{M}}_{m, k}^l$. Since the forward pass relies on exact hard selection, the structural parameter penalty $\mathcal{R}(\overline{\mathcal{M}})$ is deterministic and inherently encapsulates the coupled matrix constraints:
\begin{equation}
\begin{aligned}
\mathcal{R}(\overline{\mathcal{M}}) = \sum_{l=1}^L \sum_{k=1}^7 \Big[ & 4 \cdot (\overline{\lambda}_{\text{MSA}})_k^l \|\overline{\mathbf{M}}_{\text{MSA}, k}^l\|_0 \\
&+ 2 \cdot (\overline{\lambda}_{\text{FFN}})_k^l \|\overline{\mathbf{M}}_{\text{FFN}, k}^l\|_0 \Big].
\end{aligned}
\end{equation}

\textbf{Inference (Pruning).} 
During inference, stochastic sampling is disabled. We deterministically retain the single optimal candidate index $(k_m^*)^l$ corresponding to the highest architecture logit for each module $m$ in transformer layer $l$:
\begin{equation}
(k_m^*)^l = \operatorname*{argmax}\limits_{k} (\alpha_m)_k^l.
\end{equation}
The original weights are then \hnxu{strictly} masked exclusively by the optimal hard module-level mask $\overline{\mathbf{M}}_{m, (k_m^*)^l}^l$.

\subsection{Sparsity-Aware Self-Pinching Gates (Ours)} \label{SASPG}

The proposed \textbf{Sparsity-Aware Self-Pinching Gates (SASPG)} is also one of the compression methods using joint pruning and training. 
The core concept of SASPG is to leverage the weights already being learned to construct the mask using only one additional learnable threshold per pruned module. 
SASPG facilitates the flexible allocation of layer-wise sparsity across different layers based on their specific sensitivities, ultimately guiding the model to reach a strict size.

\subsubsection{Node-level Unstructured Pruning} \label{unstr}

\begin{figure*}[htbp]
    \centering
    \includegraphics[scale=0.5]{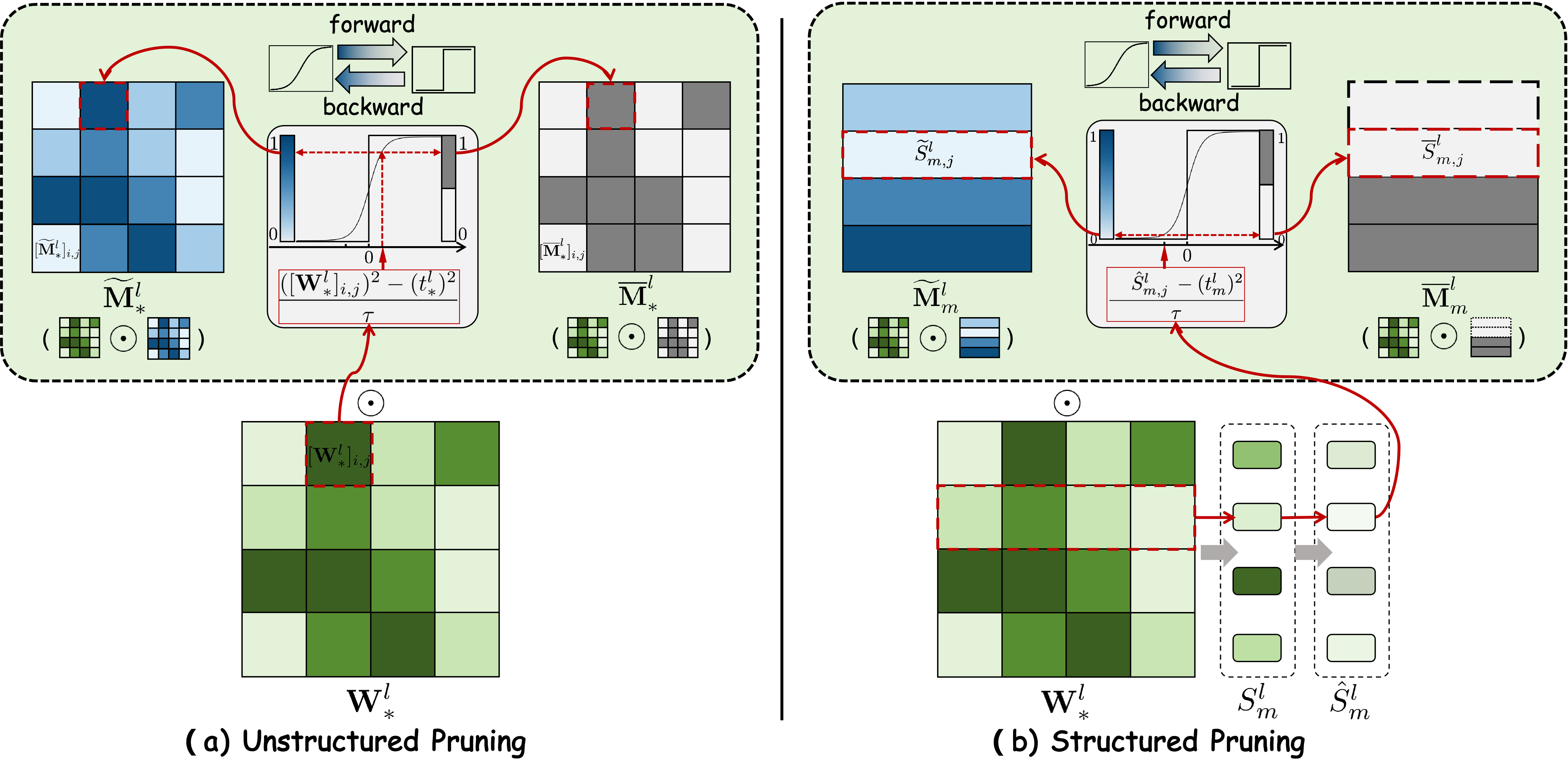}
    \caption{\textbf{Illustration of Sparsity-Aware Self-Pinching Gates (SASPG).} (a) \textbf{Unstructured Pruning}: Soft mask elements $[\widetilde{\mathbf{M}}_{*}^{l}]_{i, j}$ are derived by comparing weight magnitudes $|[\mathbf{W}_{*}^{l}]_{i, j}|$ against matrix-specific thresholds $|t_{*}^{l}|$. (b) \textbf{Structured Pruning}: \syncfix{Normalized module scores $\hat{S}_{m, j}^{l}$ are compared with the effective module threshold $(t_m^l)^2$ to construct the soft module-level mask $\widetilde{\mathbf{M}}_{m}^{l}$}, intrinsically ensuring structural consistency across all coupled matrices. Both paradigms utilize the \textbf{STE} for end-to-end training: forward passes apply the binary hard masks (e.g., $\overline{\mathbf{M}}_{*}^{l}$ or $\overline{\mathbf{M}}_{m}^{l}$), while backward passes compute gradients via the differentiable soft masks.}
    \label{fig3}
\end{figure*}

In this fine-grained approach as illustrated in Fig.~\ref{fig3} (a), we introduce a set of learnable thresholds $\mathcal{T}$ for the entire network. Recall our unified notation where $* \in \mathcal{P}$ denotes the specific weight matrix type across the network. Each individual weight matrix $\mathbf{W}_{*}^{l} \in \mathcal{W}^l$ is assigned its own independent scalar threshold $t_{*}^{l} \in \mathcal{T}$. 

For any scalar element within a specific weight matrix $\mathbf{W}_{*}^{l}$, denoted by its coordinates $(i, j)$, the corresponding soft mask value $[\widetilde{\mathbf{M}}_{*}^{l}]_{i, j}$ is computed via a smooth step-like function:
\begin{equation}
[\widetilde{\mathbf{M}}_{*}^{l}]_{i, j} = \sigma\left( \frac{([\mathbf{W}_{*}^{l}]_{i, j})^2 - (t_{*}^{l})^2}{\tau} \right),
\end{equation}
where $\sigma(\cdot)$ is the sigmoid function, $t_{*}^{l}$ is the matrix-specific threshold, and $\tau$ is a positive temperature hyperparameter controlling the sharpness of the gate.

The Straight-Through Estimator (STE)~\cite{bengio2013estimatingSTE} is adopted to enable end-to-end training:
\textbf{1) Forward Propagation:} The hard mask matrix $\overline{\mathbf{M}}_{*}^{l}$ is obtained by rounding the soft mask $\widetilde{\mathbf{M}}_{*}^{l}$. This effectively retains only the parameters in $\mathbf{W}_{*}^{l}$ whose absolute magnitudes exceed the current threshold $|t_{*}^{l}|$, while zeroing out the rest.
\textbf{2) Backward Propagation:} The gradients are computed entirely based on the differentiable soft mask values $\widetilde{\mathbf{M}}_{*}^{l}$ and are back-propagated to seamlessly update both the model weights $\mathcal{W}$ and the thresholds $\mathcal{T}$.

The overall training objective simultaneously minimizes the ASR-specific loss $\mathcal{L}_{\text{ctc}}$ and the structural penalty:
\begin{equation}
\label{eq:saspg loss}
\mathcal{L}_{\text{total}} = \mathcal{L}_{\text{ctc}} + \eta \cdot \mathcal{R}(\overline{\mathcal{M}}).
\end{equation}
$\mathcal{R}(\overline{\mathcal{M}})$ denotes the global retained parameter count exactly as defined in Section~\ref{sec:sparsity}. The configuration of $\eta$ seamlessly follows the Dynamic $\eta$-based Penalty (Eq.~\ref{eta}).

\subsubsection{Channel-level Structured Pruning}
As illustrated in Fig.~\ref{fig3} (b), we first perform 0-1 normalization on the unit importance scores (Eqs.~\ref{score_att}-\ref{score_ffn}) of the structural units to scale their values into the range \([0, 1]\). Let $\mathcal{S}_{m}^{l}$ denote the set of importance scores within module $m \in \{\text{MSA}, \text{FFN}\}$ of transformer layer $l$. The normalized score \(\hat{S}_{m, j}^{l}\) for the $j$-th structural unit is computed as:
\begin{equation}
\hat{S}_{m, j}^{l} = \frac{S_{m, j}^{l} - \min(\mathcal{S}_{m}^{l})}{\max(\mathcal{S}_{m}^{l}) - \min(\mathcal{S}_{m}^{l})}.
\end{equation}
Here, \(\min(\mathcal{S}_{m}^{l})\) and \(\max(\mathcal{S}_{m}^{l})\) denote the minimum and maximum score values within the corresponding module. To avoid division by zero, we set \(\hat{S}_{m, j}^{l} = 1 \) if all elements in \(\mathcal{S}_{m}^{l}\) are identical.

Based on the normalized scores, we generate a unified scalar soft gate $\widetilde{S}_{m, j}^{l}$ for the $j$-th structural unit by comparing it against a module-specific learnable threshold $t_m^l \in \mathcal{T}$:
\begin{equation}
\label{eq:struct_soft_gate}
\widetilde{S}_{m, j}^{l} = \sigma\left( \frac{\hat{S}_{m, j}^{l} - (t_m^l)^2}{\tau} \right).
\end{equation}

Crucially, these scalar soft gates are directly broadcast to their corresponding channel blocks to construct the soft module-level mask $\widetilde{\mathbf{M}}_{m}^{l}$. 

During the fine-tuning stage, we employ the Straight-Through Estimator (STE)~\cite{bengio2013estimatingSTE} to enable end-to-end training:
\textbf{1) Forward Propagation:} The hard module-level mask $\overline{\mathbf{M}}_{m}^{l}$ is obtained by element-wise rounding of the soft module mask $\widetilde{\mathbf{M}}_{m}^{l}$. This effectively dictates that a structural unit is retained exclusively if its normalized importance score exceeds the squared module threshold $(t_m^l)^2$, inherently guaranteeing structural integrity and physical hardware acceleration.
\textbf{2) Backward Propagation:} The gradients are computed entirely based on the differentiable soft module-level mask $\widetilde{\mathbf{M}}_{m}^{l}$ and are back-propagated to update both the original model weights $\mathcal{W}$ and the module thresholds $\mathcal{T}$.
The overall training objective for this structured variant remains fully consistent with the $\mathcal{L}_{\text{total}}$ (Eq.~\ref{eq:saspg loss}).
\revision{For SASPG with the augmented Lagrangian (AL), we use $\mathcal{L}_{\mathrm{ctc}}+\lambda_1(R_c-R_t)+\lambda_2(R_c-R_t)^2$, with the multipliers optimized as in Eq.~\ref{eq:lagrangian}.}

\section{Task-agnostic Foundation Model Pruning}
\label{sec5}

To elevate our pruning from task-specific ASR to universal SSL foundation models, we evaluate on the SUPERB benchmark~\cite{superb}. This versatile deployment requires replacing the CTC loss with a task-agnostic distillation objective. Integrating the established strategy~\cite{peng2023dphubert}, we execute \textbf{Joint Distillation and Pruning} to yield a robustly compressed upstream model fully primed for downstream fine-tuning.

Specifically, suppose that the teacher has $L^{tea}$ transformer layers with hidden size $d^{tea}$ and the student has $L^{stu}$ layers with hidden size $d^{stu}$. Let $\mathbf{X}_{l}^{tea} \in \mathbb{R}^{T \times d^{tea}}$ and $\mathbf{X}_{l}^{stu} \in \mathbb{R}^{T \times d^{stu}}$ be the output sequences of the $l$-th transformer layer of the teacher and student, respectively, where $T$ is the sequence length. To bridge the dimension mismatch, a linear projection $\mathbf{W}_{l}^{proj} \in \mathbb{R}^{d^{stu} \times d^{tea}}$ is applied to the student's output, denoted as $\hat{\mathbf{X}}_{l}^{stu} = \mathbf{X}_{l}^{stu} \mathbf{W}_{l}^{proj}$. The total distillation loss is defined as:
\begin{equation}
\mathcal{L}_{dis} = \sum_{l \in \mathcal{S}} \mathcal{L}_{l}, \label{eq:L_dis}
\end{equation}
where $\mathcal{S}$ is the set of layers chosen for matching. We set $\mathcal{S} = \{0, 4, 8, 12\}$ for base models (e.g., HuBERT-Base) and $\mathcal{S} = \{0, 8, 16, 24\}$ for large models (e.g., HuBERT-Large), with the $0$-th layer representing the CNN feature extractor output. Following \cite{peng2023dphubert, distilhubert}, the layer-wise loss $\mathcal{L}_{l}$ is a weighted combination of $L_1$ distance and cosine similarity:
\begin{equation}
\mathcal{L}_{l} = \frac{1}{T} \sum_{t=1}^{T} \left[ \frac{1}{d^{tea}} \|\mathbf{x}_{l,t}^{tea} - \hat{\mathbf{x}}_{l,t}^{stu}\|_1 - \log \sigma \left( \cos(\mathbf{x}_{l,t}^{tea}, \hat{\mathbf{x}}_{l,t}^{stu}) \right) \right],
\end{equation}
where $\mathbf{x}_{l,t}$ denotes the feature vector at time step $t$, $\sigma(\cdot)$ is the sigmoid function, and $\cos(\cdot, \cdot)$ represents the cosine similarity. Minimizing $\mathcal{L}_{l}$ simultaneously minimizes the $L_1$ distance and maximizes the cosine similarity between the hidden representations. 

\syncfix{For task-agnostic pruning, $R_c$ denotes the reduction in total foundation-model parameter count relative to the teacher.} The theoretical training objective of joint distillation and pruning can be expressed as:
\begin{equation}
\label{eq:joint_opt}
 \begin{aligned}
&\min_{\mathcal{W}, \Theta} \left[ \frac{1}{D}\sum_{k=1}^{D}\mathcal{L}_{dis}\left(f^{tea}(x_{k}), f^{stu}(x_{k}; \mathcal{W} \odot \overline{\mathcal{M}})\right) \right]\\
&
\quad \text{s.t.} \quad R_c = R_t 
\end{aligned}
\end{equation}
where $\mathcal{W} \odot \overline{\mathcal{M}}$ denotes the effective pruned weights, $\Theta$ encapsulates the method-specific pruning parameters (e.g., the thresholds $\mathcal{T}$ in our SASPG), and $R_t$ is the pre-specified target sparsity ratio. 

Different from the Dynamic $\eta$-based Penalty mechanism used in Section~\ref{ASR section}, the constrained 
objective (Eq.~\ref{eq:joint_opt}) can be converted into a minimax problem using the \syncfix{augmented Lagrangian} method~\cite{wang2020structured,peng2023dphubert}:
{\color{black}
\begin{equation}
\label{eq:lagrangian}
\begin{aligned}
\max_{\lambda_{1}, \lambda_{2}} \min_{\mathcal{W}, \Theta} &\left\{
    \frac{1}{D}\sum_{k=1}^{D}\mathcal{L}_{dis}\left(f^{tea}(x_{k}), f^{stu}(x_{k}; \mathcal{W} \odot \overline{\mathcal{M}})\right)\right.
     \\
&\left.{}+ \lambda_{1}(R_c - R_t)
    + \lambda_{2}(R_c - R_t)^{2}\right\}.
\end{aligned}
\end{equation}
}
where $\lambda_1, \lambda_2 \in \mathbb{R}$ are learnable Lagrange multipliers. \syncfix{These additional terms penalize deviations of the student model size from the target $R_t$.}

\subsection{Uniform Magnitude-based Pruning} 
As detailed in Section~\ref{ump}, the model is initially compressed to the target sparsity ($R_c = R_t$) via one-shot UMP. To subsequently recover representation quality, this framework explicitly replaces the downstream ASR objective $\mathcal{L}_{ctc}$ with the task-agnostic distillation loss $\mathcal{L}_{dis}$ (\syncfix{Eq.~\ref{eq:L_dis}}), driving a dedicated distillation stage prior to downstream fine-tuning.

\subsection{Coarse-Grained Channel-Level DPHuBERT}

\syncfix{NASP fixes a magnitude-based ordering of structural units and restricts the search to seven nested candidate masks, preventing the exploration of arbitrary combinations of retained units.} In contrast, the $L_0$-based regularization~\cite{L0reg} utilizes an impartial global sampling mechanism to select individual units dynamically. As the prevailing standard in speech SSL model compression~\cite{wang2023task, jiang2023accurate,peng2023dphubert}, we replace the NAS method with this robust $L_0$-based approach to ensure a rigorous comparison, hereafter referring to it exclusively as coarse-grained channel-level DPHuBERT~\cite{peng2023dphubert}.

To globally control \hnxu{structural} components, the discrete 1D binary masks for attention heads and FFN dimensions are relaxed to follow the differentiable Hard Concrete distribution $q_{\bm{\alpha}}$~\cite{L0reg, maddison2022concrete}. This distribution is parameterized by a set of learnable architecture parameters $\bm{\alpha} = \{\alpha_{m,j}^l\}$---which instantiates the method-specific pruning parameters $\Theta$ (Eq.~\ref{eq:joint_opt}) for this structural setting---where $\alpha_{m, j}^l$ specifically controls the $j$-th unit of module $m \in \{\text{MSA}, \text{FFN}\}$ in transformer layer $l$.

\textbf{Training.} \syncfix{A continuous variable $s_{m,j}^l$ is sampled and rectified into a Hard Concrete gate $z_{m,j}^l\in[0,1]$ for the $j$-th unit of module $m$ in transformer layer $l$:}
\begin{align}
&u \sim \mathcal{U}(0,1), \quad s_{m,j}^l = \sigma\left( \frac{1}{T} \left( \log \left(\frac{u}{1-u}\right) + \log \alpha_{m,j}^l \right) \right), \\
&\bar{s}_{m,j}^l = s_{m,j}^l \cdot (\zeta - \gamma) + \gamma, \quad z_{m,j}^l = \min(1, \max(0, \bar{s}_{m,j}^l)),
\end{align}
where $\sigma(\cdot)$ is the sigmoid function, $T$ is the temperature, and $\zeta, \gamma$ are the stretch limits. 

\syncfix{During training, the gates $z_{m,j}^l$ are broadcast to construct continuous module-level masks $\widetilde{\mathbf{M}}_{m}^{l}$, which are applied to the corresponding weight matrices.} Let $N_m$ and $c_m$ denote the total number and per-unit parameter count of structural units in module $m$. The expected retained parameter count is computed in a closed form:
{\color{black}
\begin{equation}
\label{eq:expected_penalty}
\begin{aligned}
&\mathbb{E}_{\mathbf z \sim q_{\bm{\alpha}}}\!\left[
\sum_{l=1}^L \sum_{m \in \{\text{MSA}, \text{FFN}\}} c_m
\sum_{j=1}^{N_m} \mathbb{I}[z_{m,j}^l>0]\right] \\
&\quad=\sum_{l=1}^L \sum_{m \in \{\text{MSA}, \text{FFN}\}} c_m \sum_{j=1}^{N_m} \sigma\left( \log \alpha_{m, j}^l - T \log (-\gamma/\zeta) \right).
\end{aligned}
\end{equation}
}
\syncfix{Here, $\mathbf z$ denotes all sampled Hard Concrete gates.}
\textbf{Inference (Pruning).} \syncfix{Removing the sampling noise $u$, a deterministic gate $\hat{z}_{m,j}^l\in[0,1]$ is computed directly from $\alpha_{m,j}^l$:}
\begin{equation}
\hat{z}_{m,j}^l = \min\left(1, \max\left(0, \sigma(\log \alpha_{m,j}^l)(\zeta - \gamma) + \gamma\right)\right).
\end{equation}
\syncfix{At export, the binary retention decision is $m_{m,j}^l=\mathbb{I}[\hat{z}_{m,j}^l>0]$; units with $m_{m,j}^l=0$ are physically pruned.}

\subsection{Sparsity-Aware Self-Pinching Gates (Ours)} \label{sec:saspg_distill}

All other implementation details of our method remain consistent with those described in Section~\ref{SASPG}. In this task-agnostic context, we adapt the overall training objective (Eq.~\ref{eq:saspg loss}) by substituting the ASR-specific CTC loss with the distillation loss $\mathcal{L}_{dis}$ (Eq.~\ref{eq:L_dis}). Furthermore, to align with the joint distillation framework and eliminate the need for tedious hyperparameter tuning, we replace the Dynamic $\eta$-based Penalty with the augmented Lagrangian approach (Eq.~\ref{eq:lagrangian}).

Consequently, the reformulated total objective function for our SASPG in the distillation phase is defined as:
\begin{equation}
\label{eq:saspg_distill_loss}
\begin{aligned}
\mathcal{L}_{\text{total}} = \mathcal{L}_{dis} &+ \lambda_{1} \cdot (R_c - R_{t}) + \lambda_{2} \cdot (R_c - R_{t})^{2}.
\end{aligned}
\end{equation}

\section{Experiments}
\label{sec6}

\subsection{Experimental Setup}

\textbf{Datasets, Models, and Evaluation.} 
1) \textbf{ASR-Specific Pruning:} We evaluate the {wav2vec2.0-Base} and {HuBERT-Large} architectures on the LibriSpeech~\cite{panayotov2015librispeech} 100-hour and 960-hour subsets to investigate the impact of training data scale. Performance is measured via Word Error Rate (WER) across the \texttt{dev-clean}, \texttt{dev-other}, \texttt{test-clean}, and \texttt{test-other} subsets. 
2) \textbf{Task-Agnostic Pruning via Distillation:} We expand our evaluation to include {HuBERT-Base} and {WavLM-Base+}. Following SSL model compression on the LibriSpeech datasets, the models are fine-tuned across the 10 downstream tasks of the SUPERB benchmark~\cite{superb}. We strictly adhere to default SUPERB configurations, except for Speaker Identification (SID), which utilizes a specifically tuned learning rate of $5 \times 10^{-3}$, following \cite{peng2023dphubert}.

\textbf{Implementation Details.} 
\textbf{1) ASR-Specific Pruning:} Node-level/channel-level SASPG is evaluated against offline Uniform Magnitude-based Pruning (UMP) and coarse-grained NAS-based Channel-wise Pruning (NASP). 
To establish the uncompressed baselines as our starting points, the \textit{wav2vec2.0-Base} baseline is directly adopted from \texttt{wav2vec2-base-100h}\footnote{\url{https://huggingface.co/facebook/wav2vec2-base-100h}}. For \textit{HuBERT-Large}, the baseline is obtained by first fine-tuning the pre-trained \texttt{HuBERT-large-ll60k}\footnote{\url{https://huggingface.co/facebook/HuBERT-large-ll60k}} for 20 epochs \syncfix{on LibriSpeech 100h}. Throughout all ASR-specific stages---including this baseline preparation and the subsequent pruning process---all models are optimized using AdamW with a batch size of 32 and a peak learning rate of $2 \times 10^{-4}$, employing a 10\% linear warmup followed by linear decay.
\syncfix{For LibriSpeech 100h, SASPG and NASP jointly prune and fine-tune \textit{wav2vec2.0-Base} for 30 epochs (26,760 updates) and \textit{HuBERT-Large} for 10 epochs (8,910 updates). The LibriSpeech 960h SASPG and NASP experiments use the same model-specific update budgets as their 100h counterparts, namely 26,760 and 8,910 updates, respectively. UMP first prunes each model to the target sparsity and then fine-tunes it under the corresponding baseline protocol.} For SASPG, the learnable \hnxu{thresholds} $t \in \mathcal{T}$ are initialized to $1 \times 10^{-5}$, and the temperature $\tau$ is cosine-annealed from 0.5 to 0.01.
The penalty factor $\eta$ is empirically set to $2 \times 10^{-5}$ ({wav2vec2.0-Base}) and $1 \times 10^{-6}$ ({HuBERT-Large}) for target sparsities $<65\%$, and explicitly increased to $3 \times 10^{-5}$ and $2 \times 10^{-6}$ for sparsities $\ge 65\%$. For coarse-grained channel-level NASP, $\eta$ is set to $4 \times 10^{-5}$ and $3 \times 10^{-5}$ for sparsities $<75\%$, scaling up to $2 \times 10^{-4}$ and $5 \times 10^{-5}$ for sparsities $\ge 75\%$.
\revision{For SASPG/AL, the multiplier learning rate is fixed at $2\times10^{-2}$.
For both DP and AL, we regard the target sparsity as reached when the final measured sparsity deviates from the target by no more than one percentage point.}
\textbf{2) Task-Agnostic Pruning via Distillation:} Following the established joint pruning protocol~\cite{peng2023dphubert}, we exclusively prune the Encoder transformer layers utilizing 640 seconds of audio per mini-batch. We compare SASPG against UMP and coarse-grained channel-level DPHuBERT in two stages. \textbf{Stage 1 (Joint Pruning and Distillation):} Training runs for 50k steps with a 30\% linear warmup. 
The peak learning rates are uniformly set to $2 \times 10^{-4}$ for the shared main parameters ($\mathcal{W}$), $2 \times 10^{-2}$ for the common auxiliary parameters ($\{\lambda_1, \lambda_2\}$) and the unique architecture parameter $\bm{\alpha}$ in DPHuBERT, and $1 \times 10^{-4}$ for the learnable thresholds $\mathcal{T}$ in SASPG. \syncfix{The other threshold settings are as previously described.} The target sparsity is linearly increased over the first 5k steps.
\textbf{Stage 2 (Further Distillation):} The compressed architectures are further distilled for 25k steps using a uniform peak learning rate of $1 \times 10^{-4}$ and a 20\% linear warmup. For fair comparison, UMP is pre-pruned prior to any training and then fine-tuned for 75k steps (peak learning rate $2 \times 10^{-4}$, 20\% warmup).

\subsection{Analysis of Parameter Efficiency}
Table~\ref{tab:param_feature} compares pruning consistency and parameter overheads. Compared to offline UMP, SASPG guarantees consistency between model pruning and parameter update. Furthermore, unlike coarse-grained channel-level NASP and DPHuBERT, it flexibly extends to fine-grained unstructured pruning without introducing massive auxiliary parameter overheads.
\begin{table}[htbp]
\caption{Comparison of key properties across distinct pruning methods and granularities. For SSL Base / Large models, the number of transformer layers is 12 / 24, the hidden dimension is 768 / 1024, the number of attention heads is 12 / 16, and the number of intermediate neurons is 3072 / 4096, respectively.}
\centering
\setlength\tabcolsep{3pt}
\renewcommand{\arraystretch}{1.5}
\resizebox{\linewidth}{!}{
\begin{tabular}{c|c|c|c|c|c}
\hline
\syncfix{\multirow{2}{*}{ID}} & \multirow{2}{*}{Model} & \multirow{2}{*}{\makecell{Pruning \\ Method}} & \multirow{2}{*}{\makecell{Consistency between \\ Model Pruning and \\ Parameter Update}} & \multirow{2}{*}{\makecell{Pruning \\ Granularity}} & \multirow{2}{*}{\makecell{\# Pruning-specific\\Parameters}} \\
&&&&&\\
\hline
\syncfix{1} & \multirow{6}{*}{\makecell{wav2vec2.0 Base / \\ WavLM Base+ / \\ HuBERT Base}} & \multirow{2}{*}{Magnitude} & \multirow{2}{*}{$\times$} & Node & \multirow{2}{*}{0} \\ 
\cline{5-5}
\syncfix{2} & & & & Channel & \\ 
\cline{3-6}
\syncfix{3} & & NASP & $\checkmark$ & Channel & \makecell{$2\times7\times12$ \\ $=168$} \\
\cline{3-6}
\syncfix{4} & & DPHuBERT & $\checkmark$ & Channel & \makecell{$(12+3072)\times12$ \\ $=37008$} \\
\cline{3-6}
\syncfix{5} & & \multirow{2}{*}{\makecell{SASPG\\(Ours)}} & \multirow{2}{*}{$\boldsymbol{\checkmark}$} & Node & $6\times12=72$ \\
\cline{5-6}
\syncfix{6} & & & & Channel & $2 \times 12=24$ \\
\hline
\syncfix{7} & \multirow{6}{*}{HuBERT Large} & \multirow{2}{*}{Magnitude} & \multirow{2}{*}{$\times$} & Node & \multirow{2}{*}{0}  \\ 
\cline{5-5}
\syncfix{8} & & & & Channel & \\
\cline{3-6}
\syncfix{9} & & NASP & $\checkmark$ & Channel & \makecell{$2\times7\times24$ \\ $=336$} \\
\cline{3-6}
\syncfix{10} & & DPHuBERT & $\checkmark$ & Channel & \makecell{$(16+4096)\times24$ \\ $=98688$} \\
\cline{3-6}
\syncfix{11} & & \multirow{2}{*}{\makecell{SASPG\\(Ours)}} & \multirow{2}{*}{$\boldsymbol{\checkmark}$} & Node & $6\times24=144$ \\
\cline{5-6}
\syncfix{12} & & & & Channel & $2\times24=48$ \\
\hline
\end{tabular} }
\label{tab:param_feature}
\end{table}

\subsection{Performance Analysis on the ASR-Specific Pruning}
\label{subsec:asr_results}

\begin{table*}[htbp]
    \caption{ASR-specific pruning results for wav2vec2.0-Base and HuBERT-Large. Statistical significance is evaluated via MAPSSWE~\cite{gillick1989some} ($\alpha=0.05$). $\dag$: superior to, or no significant WER increase versus, the unpruned baseline (ID~0). \textcolor{black}{Blue-colored cells}: SASPG significantly outperforms its offline UMP counterpart at the same granularity. \underline{Underlined} results: node-level pruning significantly outperforms channel-level pruning within the same method. Subscripts in parentheses denote the absolute WER reduction of SASPG over offline UMP. \revision{DP: $\eta$-based penalty tuning; AL: augmented Lagrangian.}}
    \centering
    \setlength\tabcolsep{5.6pt}  
    \resizebox{\linewidth}{!}{
    \begin{tabular}{c|c|>{\color{black}}c|c|c|c|cccc|c|cccc|c}
    \hline
    \multirow{2}{*}{ID} & \multirow{2}{*}{Method} & \multirow{2}{*}{\revision{Penalty}} & \multirow{2}{*}{Train set} & \multirow{2}{*}{Granularity} & \multirow{2}{*}{\makecell{Target\\sparsity}} &
    \multicolumn{5}{c|}{wav2vec2.0 Base (w)} & \multicolumn{5}{c}{HuBERT Large (h)} \\
    \cline{7-16}
    & & & & & & \makecell{dev-\\clean} & \makecell{dev-\\other} & \makecell{test-\\clean} & \makecell{test-\\other} & Avg. & \makecell{dev-\\clean} & \makecell{dev-\\other} & \makecell{test-\\clean} & \makecell{test-\\other} & Avg. \\
    \hline
    0 & Original & - & - & - & 0\% & 6.10 & 13.79 & 6.06 & 13.52 & 9.87 & 3.35 & 8.13 & 3.44 & 8.34 & 5.82 \\
\hline
\multicolumn{16}{c}{\textit{Method: Uniform Magnitude-based pruning (UMP)}} \\
\hline
    1 &  &  &  &  & 50\% & \underline{5.91$^\dag$} & \underline{15.69} & \underline{6.00$^\dag$} & \underline{15.39} & \underline{10.75} & \underline{3.49} & \underline{8.85} & \underline{3.64} & \underline{8.83} & \underline{6.20} \\
    2 &  &  &  &  & 60\% & \underline{6.09$^\dag$} & \underline{16.33} & \underline{6.26} & \underline{16.27} & \underline{11.24} & \underline{3.57} & \underline{9.67} & \underline{3.67} & \underline{9.41} & \underline{6.58} \\
    3 &  &  &  &  & 70\% & \underline{6.73} & \underline{18.04} & \underline{6.89} & \underline{18.00} & \underline{12.42} & \underline{4.02} & \underline{11.20} & \underline{4.23} & \underline{10.93} & \underline{7.60} \\
    4 &  &  &  &  & 80\% & \underline{8.08} & \underline{21.78} & \underline{8.64} & \underline{22.34} & \underline{15.21} & \underline{5.15} & \underline{14.35} & \underline{5.22} & \underline{14.38} & \underline{9.78} \\
    5 &  &  &  & \multirow{-5}{*}{node} & 90\% & 57.53 & 77.79 & 57.99 & 79.12 & 68.11 & \underline{10.77} & \underline{26.40} & \underline{10.99} & \underline{26.97} & \underline{18.78} \\
\hhline{~|~|~|~|-|-|----|-|----|-}
    6 &  &  &  &  & 50\% & 7.97 & 21.88 & 8.17 & 22.42 & 15.11 & 4.79 & 14.99 & 4.99 & 15.03 & 9.95 \\
    7 &  &  &  &  & 60\% & 10.36 & 27.77 & 11.01 & 28.32 & 19.37 & 5.72 & 17.40 & 5.77 & 17.82 & 11.68 \\
    8 &  &  &  &  & 70\% & 13.00 & 33.09 & 13.44 & 34.76 & 23.57 & 7.05 & 21.25 & 7.12 & 21.64 & 14.27 \\
    9 &  &  &  &  & 80\% & 17.00 & 39.67 & 17.45 & 41.54 & 28.92 & 9.67 & 27.25 & 10.06 & 28.00 & 18.75 \\
    10 &  &  & \multirow{-10}{*}{100h} & \multirow{-5}{*}{channel} & 90\% & 34.00 & 59.26 & 34.29 & 61.91 & 47.37 & 18.90 & 40.16 & 18.82 & 41.41 & 29.82 \\
\hhline{~|~|~|-|-|-|----|-|----|-}
    11 &  &  &  &  & 50\% & \underline{4.45$^\dag$} & \underline{11.13$^\dag$} & \underline{4.69$^\dag$} & \underline{10.97$^\dag$} & \underline{7.81$^\dag$} & \underline{2.77$^\dag$} & \underline{6.22$^\dag$} & \underline{2.87$^\dag$} & \underline{6.27$^\dag$} & \underline{4.53$^\dag$} \\
    12 &  &  &  &  & 60\% & \underline{4.78$^\dag$} & \underline{11.77$^\dag$} & \underline{4.95$^\dag$} & \underline{11.73$^\dag$} & \underline{8.31$^\dag$} & \underline{3.03$^\dag$} & \underline{6.83$^\dag$} & \underline{3.06$^\dag$} & \underline{6.89$^\dag$} & \underline{4.95$^\dag$} \\
    13 &  &  &  &  & 70\% & \underline{5.44$^\dag$} & \underline{12.91$^\dag$} & \underline{5.71$^\dag$} & \underline{12.92$^\dag$} & \underline{9.25$^\dag$} & \underline{3.42$^{\dag}$} & \underline{7.95$^{\dag}$} & \underline{3.51$^{\dag}$} & \underline{8.04$^\dag$} & \underline{5.73$^\dag$} \\
    14 &  &  &  &  & 80\% & \underline{7.21} & \underline{15.98} & \underline{7.44} & \underline{16.34} & \underline{11.74} & \underline{4.74} & \underline{10.77} & \underline{4.63} & \underline{11.04} & \underline{7.80} \\
    15 &  &  &  & \multirow{-5}{*}{node} & 90\% & 48.44 & 64.08 & 48.47 & 65.44 & 56.61 & \underline{10.69} & \underline{21.83} & \underline{10.99} & \underline{22.60} & \underline{16.53} \\
\hhline{~|~|~|~|-|-|----|-|----|-}
    16 &  &  &  &  & 50\% & 6.11$^\dag$ & 14.58 & 6.37 & 14.75 & 10.45 & 4.02 & 9.72 & 4.10 & 9.85 & 6.92 \\
    17 &  &  &  &  & 60\% & 8.16 & 18.44 & 8.45 & 18.71 & 13.44 & 4.91 & 12.06 & 4.97 & 12.17 & 8.53 \\
    18 &  &  &  &  & 70\% & 10.61 & 22.35 & 11.17 & 23.24 & 16.84 & 6.47 & 15.12 & 6.39 & 15.57 & 10.89 \\
    19 &  &  &  &  & 80\% & 14.42 & 28.07 & 14.55 & 29.09 & 21.53 & 9.32 & 20.83 & 9.39 & 21.56 & 15.28 \\
    20 & \multirow{-20}{*}{UMP} & \multirow{-20}{*}{-} & \multirow{-10}{*}{960h} & \multirow{-5}{*}{channel} & 90\% & 30.66 & 48.28 & 30.83 & 49.81 & 39.90 & 19.46 & 34.94 & 19.13 & 35.72 & 27.31 \\
\hline
\multicolumn{16}{c}{\textit{Method: Coarse-grained Channel-level NAS-based Pruning (NASP)}} \\
\hline
    21 &  &  &  &  & 50\% & 9.28 & 24.62 & 9.67 & 25.48 & 17.26 & 7.41 & 20.49 & 7.55 & 21.24 & 14.17 \\
    22 &  &  &  &  & 60\% & 11.79 & 31.12 & 12.11 & 32.44 & 21.87 & 9.75 & 26.23 & 9.80 & 27.23 & 18.25 \\
    23 &  &  &  &  & 70\% & 16.40 & 37.37 & 16.83 & 39.00 & 27.40 & 15.10 & 37.32 & 15.00 & 39.33 & 26.69 \\
    24 &  &  &  &  & 80\% & 24.51 & 50.20 & 24.57 & 52.42 & 37.93 & 26.29 & 49.09 & 26.47 & 50.41 & 38.07 \\
    25 &  &  & \multirow{-5}{*}{100h} & \multirow{-5}{*}{channel} & 90\% & 55.34 & 75.34 & 55.30 & 76.74 & 65.68 & 73.02 & 86.43 & 71.91 & 87.41 & 79.69 \\
\hhline{~|~|~|-|-|-|----|-|----|-}
    26 &  &  &  &  & 50\% & 8.60 & 18.74 & 8.75 & 19.62 & 13.93 & 7.21 & 16.30 & 7.45 & 17.19 & 12.04 \\
    27 &  &  &  &  & 60\% & 9.84 & 21.14 & 10.38 & 22.43 & 15.95 & 9.78 & 21.40 & 9.79 & 22.29 & 15.82 \\
    28 &  &  &  &  & 70\% & 14.14 & 28.70 & 14.38 & 29.71 & 21.73 & 15.21 & 31.03 & 15.02 & 32.98 & 23.56 \\
    29 &  &  &  &  & 80\% & 23.26 & 40.58 & 23.39 & 42.22 & 32.36 & 25.06 & 42.75 & 24.78 & 44.74 & 34.33 \\
    30 & \multirow{-10}{*}{NASP} & \multirow{-10}{*}{\dynamicpenalty} & \multirow{-5}{*}{960h} & \multirow{-5}{*}{channel} & 90\% & 54.43 & 71.24 & 54.30 & 72.06 & 63.01 & 69.44 & 82.51 & 68.21 & 84.40 & 76.14 \\
\hline
\multicolumn{16}{c}{\textit{Method: Sparsity-Aware Self-Pinching Gate (SASPG)}} \\
\hline
    31 &  &  &  &  & 50\% & \underline{5.97$^{\dag}$} & \underline{16.09} & \underline{6.00$^{\dag}$} & \underline{16.06} & \underline{11.03} & \underline{3.45$^\dag$} & \underline{9.14} & \cellcolor{cyan!20}\underline{3.51$^\dag$} & \underline{8.99} & \underline{6.27} \\
    32 &  &  &  &  & 60\% & \underline{6.06$^{\dag}$} & \underline{16.52} & \cellcolor{cyan!20}\underline{6.07$^{\dag}$} & \underline{16.08} & \underline{11.18} & \underline{3.54} & \cellcolor{cyan!20}\underline{9.44} & \cellcolor{cyan!20}\underline{3.54$^\dag$} & \underline{9.33} & \cellcolor{cyan!20}\underline{6.46} \\
    33 &  &  &  &  & 70\% & \cellcolor{cyan!20}\underline{6.21$^{\dag}$} & \cellcolor{cyan!20}\underline{17.17} & \cellcolor{cyan!20}\underline{6.45} & \cellcolor{cyan!20}\underline{17.46} & \cellcolor{cyan!20}\underline{11.82} & \cellcolor{cyan!20}\underline{3.76} & \cellcolor{cyan!20}\underline{10.27} & \cellcolor{cyan!20}\underline{3.76} & \cellcolor{cyan!20}\underline{10.15} & \cellcolor{cyan!20}\underline{6.99} \\
    34 &  &  &  &  & 80\% & \cellcolor{cyan!20}\underline{6.71} & \cellcolor{cyan!20}\underline{18.63} & \cellcolor{cyan!20}\underline{6.95} & \cellcolor{cyan!20}\underline{19.27} & \cellcolor{cyan!20}\underline{12.89} & \cellcolor{cyan!20}\underline{4.14} & \cellcolor{cyan!20}\underline{12.39} & \cellcolor{cyan!20}\underline{4.31} & \cellcolor{cyan!20}\underline{12.34} & \cellcolor{cyan!20}\underline{8.30} \\
    35 &  &  &  & \multirow{-5}{*}{node} & 90\% & \cellcolor{cyan!20}\underline{8.22$_{(49.31)}$} & \cellcolor{cyan!20}\underline{23.36$_{(54.43)}$} & \cellcolor{cyan!20}\underline{8.63$_{(49.36)}$} & \cellcolor{cyan!20}\underline{24.30$_{(54.82)}$} & \cellcolor{cyan!20}\underline{16.13$_{(51.98)}$} & \cellcolor{cyan!20}\underline{6.11$_{(4.66)}$} & \cellcolor{cyan!20}\underline{18.91$_{(7.49)}$} & \cellcolor{cyan!20}\underline{6.37$_{(4.62)}$} & \cellcolor{cyan!20}\underline{19.29$_{(7.68)}$} & \cellcolor{cyan!20}\underline{12.67$_{(6.11)}$} \\
\hhline{~|~|~|~|-|-|----|-|----|-}
    36 &  &  &  &  & 50\% & \cellcolor{cyan!20}6.46 & \cellcolor{cyan!20}18.28 & \cellcolor{cyan!20}6.63 & \cellcolor{cyan!20}18.53 & \cellcolor{cyan!20}12.48 & \cellcolor{cyan!20}4.48 & 14.80 & \cellcolor{cyan!20}4.74 & 15.33 & 9.84 \\
    37 &  &  &  &  & 60\% & \cellcolor{cyan!20}7.03 & \cellcolor{cyan!20}19.88 & \cellcolor{cyan!20}7.23 & \cellcolor{cyan!20}20.51 & \cellcolor{cyan!20}13.66 & \cellcolor{cyan!20}5.18 & \cellcolor{cyan!20}17.05 & \cellcolor{cyan!20}5.29 & \cellcolor{cyan!20}17.41 & \cellcolor{cyan!20}11.23 \\
    38 &  &  &  &  & 70\% & \cellcolor{cyan!20}7.94 & \cellcolor{cyan!20}22.52 & \cellcolor{cyan!20}8.25 & \cellcolor{cyan!20}23.52 & \cellcolor{cyan!20}15.56 & \cellcolor{cyan!20}6.14 & \cellcolor{cyan!20}19.38 & \cellcolor{cyan!20}6.19 & \cellcolor{cyan!20}20.20 & \cellcolor{cyan!20}12.98 \\
    39 &  &  &  &  & 80\% & \cellcolor{cyan!20}10.44 & \cellcolor{cyan!20}28.73 & \cellcolor{cyan!20}11.30 & \cellcolor{cyan!20}30.10 & \cellcolor{cyan!20}20.14 & \cellcolor{cyan!20}8.23 & \cellcolor{cyan!20}24.18 & \cellcolor{cyan!20}8.25 & \cellcolor{cyan!20}25.45 & \cellcolor{cyan!20}16.53 \\
    40 &  & \multirow{-10}{*}{\dynamicpenalty} &  & \multirow{-5}{*}{channel} & 90\% & \cellcolor{cyan!20}22.05$_{(11.95)}$ & \cellcolor{cyan!20}45.44$_{(13.82)}$ & \cellcolor{cyan!20}22.55$_{(11.74)}$ & \cellcolor{cyan!20}47.52$_{(14.39)}$ & \cellcolor{cyan!20}34.39$_{(12.98)}$ & \cellcolor{cyan!20}16.72$_{(2.18)}$ & \cellcolor{cyan!20}37.49$_{(2.67)}$ & \cellcolor{cyan!20}16.69$_{(2.13)}$ & \cellcolor{cyan!20}38.51$_{(2.90)}$ & \cellcolor{cyan!20}27.35$_{(2.47)}$ \\
\hhline{~|~|-|~|-|-|----|-|----|-}
\revision{41} & \revision{} & \revision{} & \revision{} & \revision{} & \revision{50\%} & \revision{\underline{5.84$^\dag$}} & \revision{\underline{16.07}} & \revision{\underline{6.00$^\dag$}} & \revision{\underline{16.10}} & \revision{\underline{11.00}} & \revision{\underline{3.42$^\dag$}} & \revision{\underline{9.34}} & \revision{\underline{3.55}} & \revision{\underline{9.24}} & \revision{\underline{6.39}} \\
\revision{42} & \revision{} & \revision{} & \revision{} & \revision{} & \revision{60\%} & \cellcolor{cyan!20}\revision{\underline{5.79$^\dag$}} & \revision{\underline{16.22}} & \revision{\underline{6.11$^\dag$}} & \revision{\underline{16.18}} & \cellcolor{cyan!20}\revision{\underline{11.07}} & \revision{\underline{3.46$^\dag$}} & \revision{\underline{9.87}} & \revision{\underline{3.74}} & \revision{\underline{9.74}} & \revision{\underline{6.70}} \\
\revision{43} & \revision{} & \revision{} & \revision{} & \revision{} & \revision{70\%} & \cellcolor{cyan!20}\revision{\underline{6.16$^\dag$}} & \cellcolor{cyan!20}\revision{\underline{16.91}} & \cellcolor{cyan!20}\revision{\underline{6.18$^\dag$}} & \cellcolor{cyan!20}\revision{\underline{17.05}} & \cellcolor{cyan!20}\revision{\underline{11.58}} & \cellcolor{cyan!20}\revision{\underline{3.80}} & \cellcolor{cyan!20}\revision{\underline{10.16}} & \cellcolor{cyan!20}\revision{\underline{3.92}} & \cellcolor{cyan!20}\revision{\underline{10.16}} & \cellcolor{cyan!20}\revision{\underline{7.01}} \\
\revision{44} & \revision{} & \revision{} & \revision{} & \revision{} & \revision{80\%} & \cellcolor{cyan!20}\revision{\underline{6.63}} & \cellcolor{cyan!20}\revision{\underline{19.00}} & \cellcolor{cyan!20}\revision{\underline{6.96}} & \cellcolor{cyan!20}\revision{\underline{19.22}} & \cellcolor{cyan!20}\revision{\underline{12.95}} & \cellcolor{cyan!20}\revision{\underline{4.68}} & \cellcolor{cyan!20}\revision{\underline{13.81}} & \cellcolor{cyan!20}\revision{\underline{4.88}} & \cellcolor{cyan!20}\revision{\underline{13.88}} & \cellcolor{cyan!20}\revision{\underline{9.31}} \\
\revision{45} & \revision{} & \revision{} & \revision{} & \revision{\multirow{-5}{*}{node}} & \revision{90\%} & \cellcolor{cyan!20}\revision{\underline{8.67}} & \cellcolor{cyan!20}\revision{\underline{24.54}} & \cellcolor{cyan!20}\revision{\underline{9.13}} & \cellcolor{cyan!20}\revision{\underline{25.29}} & \cellcolor{cyan!20}\revision{\underline{16.91}} & \cellcolor{cyan!20}\revision{\underline{7.35}} & \cellcolor{cyan!20}\revision{\underline{21.85}} & \cellcolor{cyan!20}\revision{\underline{7.62}} & \cellcolor{cyan!20}\revision{\underline{22.63}} & \cellcolor{cyan!20}\revision{\underline{14.86}} \\
\hhline{~|~|~|~|-|-|----|-|----|-}
\revision{46} & \revision{} & \revision{} & \revision{} & \revision{} & \revision{50\%} & \revision{6.29$^\dag$} & \revision{18.11} & \revision{6.67} & \revision{18.51} & \revision{12.40} & \revision{4.77} & \cellcolor{cyan!20}\revision{14.29} & \cellcolor{cyan!20}\revision{4.78} & \cellcolor{cyan!20}\revision{14.62} & \cellcolor{cyan!20}\revision{9.61} \\
\revision{47} & \revision{} & \revision{} & \revision{} & \revision{} & \revision{60\%} & \revision{6.97} & \revision{19.65} & \revision{7.09} & \revision{20.11} & \revision{13.46} & \revision{5.65} & \revision{18.45} & \revision{5.61} & \revision{18.98} & \revision{12.17} \\
\revision{48} & \revision{} & \revision{} & \revision{} & \revision{} & \revision{70\%} & \revision{7.88} & \revision{22.53} & \revision{8.34} & \revision{23.03} & \revision{15.45} & \cellcolor{cyan!20}\revision{6.03} & \cellcolor{cyan!20}\revision{19.42} & \cellcolor{cyan!20}\revision{6.03} & \cellcolor{cyan!20}\revision{19.97} & \cellcolor{cyan!20}\revision{12.86} \\
\revision{49} & \revision{} & \revision{} & \revision{} & \revision{} & \revision{80\%} & \cellcolor{cyan!20}\revision{10.99} & \cellcolor{cyan!20}\revision{29.31} & \cellcolor{cyan!20}\revision{11.23} & \cellcolor{cyan!20}\revision{30.84} & \cellcolor{cyan!20}\revision{20.59} & \revision{10.99} & \revision{29.45} & \revision{11.14} & \revision{30.53} & \revision{20.53} \\
\revision{50} & \revision{} & \revision{\multirow{-10}{*}{\augmentedpenalty}} & \syncfix{\multirow{-20}{*}{100h}} & \revision{\multirow{-5}{*}{channel}} & \revision{90\%} & \cellcolor{cyan!20}\revision{21.18} & \cellcolor{cyan!20}\revision{45.07} & \cellcolor{cyan!20}\revision{21.97} & \cellcolor{cyan!20}\revision{47.27} & \cellcolor{cyan!20}\revision{33.87} & \revision{19.30} & \revision{40.58} & \revision{19.42} & \revision{41.98} & \revision{30.32} \\
\hhline{~|~|-|-|-|-|----|-|----|-}
    51 &  &  &  &  & 50\% & \underline{4.34$^\dag$} & \underline{11.21$^\dag$} & \underline{4.63$^\dag$} & \underline{11.02$^\dag$} & \underline{7.80$^\dag$} & \underline{2.70$^\dag$} & \cellcolor{cyan!20}\underline{6.03$^\dag$} & \cellcolor{cyan!20}\underline{2.76$^\dag$} & \underline{6.32$^\dag$} & \cellcolor{cyan!20}\underline{4.45$^\dag$} \\
    52 &  &  &  &  & 60\% & \cellcolor{cyan!20}\underline{4.54$^\dag$} & \cellcolor{cyan!20}\underline{11.34$^\dag$} & \cellcolor{cyan!20}\underline{4.77$^\dag$} & \cellcolor{cyan!20}\underline{11.30$^\dag$} & \cellcolor{cyan!20}\underline{7.99$^\dag$} & \cellcolor{cyan!20}\underline{2.89$^\dag$} & \cellcolor{cyan!20}\underline{6.39$^\dag$} & \cellcolor{cyan!20}\underline{2.90$^\dag$} & \cellcolor{cyan!20}\underline{6.56$^\dag$} & \cellcolor{cyan!20}\underline{4.69$^\dag$} \\
    53 &  &  &  &  & 70\% & \cellcolor{cyan!20}\underline{4.86$^\dag$} & \cellcolor{cyan!20}\underline{12.15$^\dag$} & \cellcolor{cyan!20}\underline{4.98$^\dag$} & \cellcolor{cyan!20}\underline{12.12$^\dag$} & \cellcolor{cyan!20}\underline{8.53$^\dag$} & \cellcolor{cyan!20}\underline{3.00$^\dag$} & \cellcolor{cyan!20}\underline{6.86$^\dag$} & \cellcolor{cyan!20}\underline{3.08$^\dag$} & \cellcolor{cyan!20}\underline{7.04$^\dag$} & \cellcolor{cyan!20}\underline{5.00$^\dag$} \\
    54 &  &  &  &  & 80\% & \cellcolor{cyan!20}\underline{5.34$^\dag$} & \cellcolor{cyan!20}\underline{12.75$^\dag$} & \cellcolor{cyan!20}\underline{5.49$^\dag$} & \cellcolor{cyan!20}\underline{12.88$^\dag$} & \cellcolor{cyan!20}\underline{9.12$^\dag$} & \cellcolor{cyan!20}\underline{3.52} & \cellcolor{cyan!20}\underline{8.48} & \cellcolor{cyan!20}\underline{3.53$^{\dag}$} & \cellcolor{cyan!20}\underline{8.46$^{\dag}$} & \cellcolor{cyan!20}\underline{6.00} \\
    55 &  &  &  & \multirow{-5}{*}{node} & 90\% & \cellcolor{cyan!20}\underline{6.87$_{(41.57)}$} & \cellcolor{cyan!20}\underline{15.83$_{(48.25)}$} & \cellcolor{cyan!20}\underline{7.09$_{(41.38)}$} & \cellcolor{cyan!20}\underline{16.31$_{(49.13)}$} & \cellcolor{cyan!20}\underline{11.53$_{(45.08)}$} & \cellcolor{cyan!20}\underline{5.16$_{(5.53)}$} & \cellcolor{cyan!20}\underline{12.49$_{(9.34)}$} & \cellcolor{cyan!20}\underline{5.22$_{(5.77)}$} & \cellcolor{cyan!20}\underline{12.82$_{(9.78)}$} & \cellcolor{cyan!20}\underline{8.92$_{(7.61)}$} \\
\hhline{~|~|~|~|-|-|----|-|----|-}
    56 &  &  &  &  & 50\% & \cellcolor{cyan!20}5.19$^\dag$ & \cellcolor{cyan!20}12.65$^\dag$ & \cellcolor{cyan!20}5.43$^\dag$ & \cellcolor{cyan!20}12.83$^\dag$ & \cellcolor{cyan!20}9.03$^\dag$ & \cellcolor{cyan!20}3.68 & \cellcolor{cyan!20}8.80 & \cellcolor{cyan!20}3.71 & \cellcolor{cyan!20}8.86 & \cellcolor{cyan!20}6.26 \\
    57 &  &  &  &  & 60\% & \cellcolor{cyan!20}5.88$^\dag$ & \cellcolor{cyan!20}14.03$^{\dag}$ & \cellcolor{cyan!20}6.21$^{\dag}$ & \cellcolor{cyan!20}14.24 & \cellcolor{cyan!20}10.09 & \cellcolor{cyan!20}4.43 & \cellcolor{cyan!20}10.70 & \cellcolor{cyan!20}4.34 & \cellcolor{cyan!20}10.93 & \cellcolor{cyan!20}7.60 \\
    58 &  &  &  &  & 70\% & \cellcolor{cyan!20}7.27 & \cellcolor{cyan!20}16.60 & \cellcolor{cyan!20}7.57 & \cellcolor{cyan!20}17.36 & \cellcolor{cyan!20}12.20 & \cellcolor{cyan!20}5.47 & \cellcolor{cyan!20}13.37 & \cellcolor{cyan!20}5.43 & \cellcolor{cyan!20}13.57 & \cellcolor{cyan!20}9.46 \\
    59 &  &  &  &  & 80\% & \cellcolor{cyan!20}10.48 & \cellcolor{cyan!20}21.85 & \cellcolor{cyan!20}10.71 & \cellcolor{cyan!20}22.48 & \cellcolor{cyan!20}16.38 & \cellcolor{cyan!20}7.97$_{(1.35)}$ & \cellcolor{cyan!20}17.91$_{(2.92)}$ & \cellcolor{cyan!20}7.84$_{(1.55)}$ & \cellcolor{cyan!20}18.60$_{(2.96)}$ & \cellcolor{cyan!20}13.08$_{(2.20)}$ \\
    60 &  & \multirow{-10}{*}{\dynamicpenalty} &  & \multirow{-5}{*}{channel} & 90\% & \cellcolor{cyan!20}21.91$_{(8.75)}$ & \cellcolor{cyan!20}37.69$_{(10.59)}$ & \cellcolor{cyan!20}22.39$_{(8.44)}$ & \cellcolor{cyan!20}38.93$_{(10.88)}$ & \cellcolor{cyan!20}30.23$_{(9.67)}$ & \cellcolor{cyan!20}17.85 & \cellcolor{cyan!20}32.44 & \cellcolor{cyan!20}17.63 & \cellcolor{cyan!20}33.53 & \cellcolor{cyan!20}25.36 \\
\hhline{~|~|-|~|-|-|----|-|----|-}
\revision{61} & \revision{} & \revision{} & \revision{} & \revision{} & \revision{50\%} & \revision{\underline{4.33$^\dag$}} & \revision{\underline{11.00$^\dag$}} & \cellcolor{cyan!20}\revision{\underline{4.38$^\dag$}} & \revision{\underline{11.03$^\dag$}} & \cellcolor{cyan!20}\revision{\underline{7.69$^\dag$}} & \revision{\underline{2.73$^\dag$}} & \revision{\underline{6.16$^\dag$}} & \revision{\underline{2.79$^\dag$}} & \revision{\underline{6.18$^\dag$}} & \cellcolor{cyan!20}\revision{\underline{4.46$^\dag$}} \\
\revision{62} & \revision{} & \revision{} & \revision{} & \revision{} & \revision{60\%} & \revision{\underline{4.44$^\dag$}} & \revision{\underline{11.15$^\dag$}} & \revision{\underline{4.69$^\dag$}} & \revision{\underline{11.25$^\dag$}} & \revision{\underline{7.88$^\dag$}} & \revision{\underline{2.87$^\dag$}} & \revision{\underline{6.23$^\dag$}} & \revision{\underline{2.86$^\dag$}} & \revision{\underline{6.51$^\dag$}} & \revision{\underline{4.62$^\dag$}} \\
\revision{63} & \revision{} & \revision{} & \revision{} & \revision{} & \revision{70\%} & \revision{\underline{4.86$^\dag$}} & \revision{\underline{12.20$^\dag$}} & \revision{\underline{5.10$^\dag$}} & \revision{\underline{12.27$^\dag$}} & \revision{\underline{8.61$^\dag$}} & \revision{\underline{3.03$^\dag$}} & \revision{\underline{7.17$^\dag$}} & \revision{\underline{3.10$^\dag$}} & \revision{\underline{7.29$^\dag$}} & \revision{\underline{5.15$^\dag$}} \\
\revision{64} & \revision{} & \revision{} & \revision{} & \revision{} & \revision{80\%} & \revision{\underline{5.22$^\dag$}} & \revision{\underline{12.72$^\dag$}} & \revision{\underline{5.66$^\dag$}} & \revision{\underline{12.96$^\dag$}} & \revision{\underline{9.14$^\dag$}} & \cellcolor{cyan!20}\revision{\underline{3.64}} & \cellcolor{cyan!20}\revision{\underline{8.71}} & \cellcolor{cyan!20}\revision{\underline{3.67}} & \cellcolor{cyan!20}\revision{\underline{8.92}} & \cellcolor{cyan!20}\revision{\underline{6.23}} \\
\revision{65} & \revision{} & \revision{} & \revision{} & \revision{\multirow{-5}{*}{node}} & \revision{90\%} & \revision{\underline{6.94}} & \revision{\underline{16.02}} & \revision{\underline{7.34}} & \revision{\underline{16.69}} & \revision{\underline{11.75}} & \revision{\underline{6.92}} & \revision{\underline{16.16}} & \revision{\underline{7.01}} & \revision{\underline{16.67}} & \revision{\underline{11.69}} \\
\hhline{~|~|~|~|-|-|----|-|----|-}
\revision{66} & \revision{} & \revision{} & \revision{} & \revision{} & \revision{50\%} & \cellcolor{cyan!20}\revision{5.31$^\dag$} & \cellcolor{cyan!20}\revision{12.75$^\dag$} & \cellcolor{cyan!20}\revision{5.57$^\dag$} & \cellcolor{cyan!20}\revision{12.94$^\dag$} & \cellcolor{cyan!20}\revision{9.14$^\dag$} & \cellcolor{cyan!20}\revision{3.43$^\dag$} & \cellcolor{cyan!20}\revision{8.26$^\dag$} & \cellcolor{cyan!20}\revision{3.49$^\dag$} & \cellcolor{cyan!20}\revision{8.56$^\dag$} & \cellcolor{cyan!20}\revision{5.94} \\
\revision{67} & \revision{} & \revision{} & \revision{} & \revision{} & \revision{60\%} & \revision{8.42} & \revision{18.75} & \revision{8.48} & \revision{19.51} & \revision{13.79} & \cellcolor{cyan!20}\revision{4.03} & \cellcolor{cyan!20}\revision{9.94} & \cellcolor{cyan!20}\revision{4.08} & \cellcolor{cyan!20}\revision{10.34} & \cellcolor{cyan!20}\revision{7.10} \\
\revision{68} & \revision{} & \revision{} & \revision{} & \revision{} & \revision{70\%} & \cellcolor{cyan!20}\revision{7.67} & \cellcolor{cyan!20}\revision{17.13} & \cellcolor{cyan!20}\revision{7.96} & \cellcolor{cyan!20}\revision{17.95} & \cellcolor{cyan!20}\revision{12.68} & \cellcolor{cyan!20}\revision{5.77} & \cellcolor{cyan!20}\revision{14.10} & \cellcolor{cyan!20}\revision{5.88} & \cellcolor{cyan!20}\revision{14.53} & \cellcolor{cyan!20}\revision{10.07} \\
\revision{69} & \revision{} & \revision{} & \revision{} & \revision{} & \revision{80\%} & \cellcolor{cyan!20}\revision{14.06} & \cellcolor{cyan!20}\revision{27.63} & \cellcolor{cyan!20}\revision{14.23} & \revision{28.79} & \cellcolor{cyan!20}\revision{21.18} & \revision{11.13} & \revision{23.66} & \revision{11.09} & \revision{24.25} & \revision{17.53} \\
\revision{70} & \syncfix{\multirow{-40}{*}{SASPG}} & \revision{\multirow{-10}{*}{\augmentedpenalty}} & \syncfix{\multirow{-20}{*}{960h}} & \revision{\multirow{-5}{*}{channel}} & \revision{90\%} & \revision{34.20} & \revision{49.90} & \revision{34.46} & \revision{50.89} & \revision{42.36} & \revision{22.01} & \revision{37.69} & \revision{21.88} & \revision{38.68} & \revision{30.06} \\

\hline

    \end{tabular} }
    \par

    \label{tab:asr}
\end{table*}

\begin{table}[!t]
\caption{\revision{SASPG ASR results at 70\% sparsity initialized from the \texttt{facebook/hubert-large-ls960-ft} checkpoint.}}
\label{tab:asr_hubert_ls960_init}
\centering
\fontsize{7}{8.4}\selectfont
\setlength{\tabcolsep}{2pt}
\renewcommand{\arraystretch}{1.1}
\begin{tabular*}{\columnwidth}{@{\extracolsep{\fill}}c|c|c|c|ccccc@{}}
\hline
\revision{ID} & \revision{\makecell{Train\\set}} & \revision{Granularity} & \revision{Penalty} & \revision{\makecell{dev-\\clean}} & \revision{\makecell{dev-\\other}} & \revision{\makecell{test-\\clean}} & \revision{\makecell{test-\\other}} & \revision{Avg.} \\
\hline
\revision{1} & \revision{100h} & \revision{node} & \revision{DP} & \revision{2.97} & \revision{7.91} & \revision{3.05} & \revision{7.67} & \revision{5.40} \\
\revision{2} & \revision{100h} & \revision{node} & \revision{AL} & \revision{3.04} & \revision{8.30} & \revision{3.10} & \revision{8.06} & \revision{5.63} \\
\revision{3} & \revision{100h} & \revision{channel} & \revision{DP} & \revision{6.37} & \revision{19.26} & \revision{6.35} & \revision{20.14} & \revision{13.03} \\
\revision{4} & \revision{100h} & \revision{channel} & \revision{AL} & \revision{5.59} & \revision{18.11} & \revision{5.70} & \revision{18.89} & \revision{12.07} \\
\hline
\revision{5} & \revision{960h} & \revision{node} & \revision{DP} & \revision{2.17} & \revision{5.79} & \revision{2.34} & \revision{5.88} & \revision{4.05} \\
\revision{6} & \revision{960h} & \revision{node} & \revision{AL} & \revision{2.64} & \revision{6.16} & \revision{2.68} & \revision{6.34} & \revision{4.46} \\
\revision{7} & \revision{960h} & \revision{channel} & \revision{DP} & \revision{5.30} & \revision{13.11} & \revision{5.26} & \revision{13.16} & \revision{9.21} \\
\revision{8} & \revision{960h} & \revision{channel} & \revision{AL} & \revision{5.85} & \revision{14.03} & \revision{5.68} & \revision{14.19} & \revision{9.94} \\
\hline
\end{tabular*}
\par\vspace{2pt}
\end{table}

The ASR-specific pruning results are summarized in {Table~\ref{tab:asr}}. Our analysis is organized around two core dimensions: performance-oriented fine-grained unstructured pruning and acceleration-oriented coarse-grained structured pruning.

\noindent\revision{The starting checkpoint in Table~\ref{tab:asr_hubert_ls960_init} differs from that of the historical HuBERT-Large results in Table~\ref{tab:asr}.}

\subsubsection{Performance-Oriented Node-level SASPG}

\textbf{i) Node-level SASPG vs. Node-level UMP.} Driven by its consistency between model pruning and parameter update, node-level SASPG consistently outperforms node-level offline UMP at sparsities of \textbf{70\%} and above. Crucially, it prevents the catastrophic collapse experienced by node-level offline UMP at extreme sparsity of \textbf{90\%}, where it delivers average absolute (relative) WER reductions of \textbf{51.98\%} (\textbf{76.32\%}) and \textbf{7.61\%} (\textbf{46.04\%}) over node-level offline UMP on wav2vec2.0-Base (ID~35w vs. ID~5w) and HuBERT-Large (ID~55h vs. ID~15h), respectively.
\textbf{ii) Node-level SASPG vs. Channel-level SASPG.} Across all evaluated sparsity levels, node-level SASPG maintains its fine-grained advantage, consistently outperforming its channel-level counterpart (\textbf{underlined results}). In contrast, node-level offline UMP on wav2vec2.0-Base (ID~5w vs. ID~10w, ID~15w vs. ID~20w) severely collapses at \textbf{90\%} sparsity, falling strictly below channel-level offline UMP.
\textbf{iii) Node-level SASPG vs. Channel-level NASP and Channel-level UMP.} Across all evaluated sparsity levels, node-level SASPG consistently improves over both coarse-grained channel-level NASP (IDs~21--30) and channel-level offline UMP baselines (IDs~6--10 and IDs~16--20).
\textbf{iv) Node-level SASPG vs. Unpruned baseline.} Pushing the boundaries of lossless compression (\textbf{$\dag$ symbol}), node-level SASPG achieves maximum lossless sparsities (\syncfix{compression ratios}) of \textbf{80\%} (\textbf{5$\times$}) and \textbf{70\%} (\textbf{3.3$\times$}) on wav2vec2.0-Base (ID~54w vs. ID~0w) and HuBERT-Large (ID~53h vs. ID~0h), respectively.

\subsubsection{Acceleration-oriented Channel-level SASPG}

\textbf{i) Channel-level SASPG vs. Channel-level NASP.} Channel-level SASPG consistently surpasses coarse-grained channel-level NASP across all evaluated subsets by dynamically bypassing its rigid mask constraints (IDs~36--40 vs. IDs~21--25; IDs~56--60 vs. IDs~26--30). In terms of pruning-specific parameters shown in {Table~\ref{tab:param_feature}}, it requires only \textbf{14.3\%} (\textbf{24}/\textbf{168} and \textbf{48}/\textbf{336}) of the parameters demanded by NASP for both models.
\textbf{ii) Channel-level SASPG vs. Channel-level UMP.} \syncfix{\revision{With DP,} channel-level SASPG outperforms channel-level offline UMP at sparsities of \textbf{60\%} and above. This advantage peaks at the extreme \textbf{90\%} sparsity, yielding maximum average absolute (relative) WER reductions of \textbf{12.98\%} (\textbf{27.40\%}) and \textbf{2.47\%} (\textbf{8.28\%}) over channel-level offline UMP on wav2vec2.0-Base (ID~40w vs. ID~10w) and HuBERT-Large (ID~40h vs. ID~10h), respectively. \revision{The added SASPG/AL results exhibit setup-dependent comparisons with UMP and are reported separately in Table~\ref{tab:asr}.}}
\textbf{iii) Channel-level SASPG vs. Unpruned baseline.} On wav2vec2.0-Base, channel-level SASPG achieves a \textbf{2$\times$} lossless \syncfix{compression ratio} (\textbf{$\dag$ symbol}) at \textbf{50\%} sparsity, with significant WER reductions across all subsets (ID~56w vs. ID~0w).

\noindent\revision{\textbf{Measured hardware performance.} Table~\ref{tab:hardware_q2} reports acoustic-model forward real-time factor (RTF), peak allocated GPU memory, and GFLOPs on an NVIDIA H800. RTF is the forward time divided by the input audio duration. At 50--90\% sparsity, channel-level SASPG/DP achieves 1.247--1.618$\times$ speedup and 30.73--56.12\% lower peak memory on wav2vec2.0-Base, and 1.289--1.868$\times$ speedup and 39.86--72.84\% lower peak memory on HuBERT-Large, relative to their dense baselines.}

\begin{table}[!t]
\caption{\revision{Acoustic-forward performance on an NVIDIA H800 with FP32, batch size 1, and 10-second inputs.} \revision{Each speedup uses the dense timing measured alongside that pruned model; the dense row shows a representative measurement.}}
\label{tab:hardware_q2}
\centering\fontsize{7}{8.4}\selectfont
\setlength{\tabcolsep}{3pt}
\renewcommand{\arraystretch}{1.1}
\begin{tabular*}{\columnwidth}{@{\extracolsep{\fill}}c|c|c|c|ccc}
\hline
\revision{ID} & \revision{\makecell{Target\\(\%)}} & \revision{Penalty} & \revision{GFLOPs} & \revision{\makecell{RTF\\$(\times10^{-3})$}} & \revision{Speedup} & \revision{\makecell{Memory\\(MiB)}} \\
\hline
\multicolumn{7}{c}{\revision{wav2vec2.0 Base}} \\
\hline
\revision{1} & \revision{0} & \revision{-} & \revision{148.16} & \revision{0.866} & \revision{1.000$\times$} & \revision{518.70} \\
\revision{2} & \revision{50} & \revision{DP} & \revision{100.79} & \revision{0.694} & \revision{1.247$\times$} & \revision{359.32} \\
\revision{3} & \revision{60} & \revision{DP} & \revision{91.24} & \revision{0.648} & \revision{1.338$\times$} & \revision{325.89} \\
\revision{4} & \revision{70} & \revision{DP} & \revision{82.37} & \revision{0.607} & \revision{1.428$\times$} & \revision{293.52} \\
\revision{5} & \revision{80} & \revision{DP} & \revision{73.18} & \revision{0.571} & \revision{1.516$\times$} & \revision{259.25} \\
\revision{6} & \revision{90} & \revision{DP} & \revision{63.80} & \revision{0.535} & \revision{1.618$\times$} & \revision{227.62} \\
\hline
\multicolumn{7}{c}{\revision{HuBERT Large}} \\
\hline
\revision{7} & \revision{0} & \revision{-} & \revision{383.89} & \revision{1.821} & \revision{1.000$\times$} & \revision{1423.78} \\
\revision{8} & \revision{50} & \revision{DP} & \revision{220.60} & \revision{1.413} & \revision{1.289$\times$} & \revision{856.22} \\
\revision{9} & \revision{60} & \revision{DP} & \revision{186.86} & \revision{1.293} & \revision{1.405$\times$} & \revision{730.43} \\
\revision{10} & \revision{70} & \revision{DP} & \revision{153.25} & \revision{1.204} & \revision{1.512$\times$} & \revision{621.31} \\
\revision{11} & \revision{80} & \revision{DP} & \revision{125.97} & \revision{1.104} & \revision{1.648$\times$} & \revision{504.66} \\
\revision{12} & \revision{90} & \revision{DP} & \revision{92.48} & \revision{0.977} & \revision{1.868$\times$} & \revision{386.68} \\
\hline
\end{tabular*}
\end{table}

\begin{table*}[htbp]
\caption{Task-agnostic joint distillation and pruning results on the SUPERB benchmark. \textbf{Bold} results: Node- or channel-level SASPG achieves the optimal performance for a given model within its respective data scale (100h/960h). \textcolor{black}{Blue-colored cells}: SASPG outperforms its offline UMP counterpart at the same granularity. $^{*}$: Channel-level SASPG outperforms its DPHuBERT counterpart. \underline{Underlined} results: Node-level pruning outperforms channel-level pruning within the same method.}
    \centering
    \setlength\tabcolsep{3pt}  
    \renewcommand{\arraystretch}{1.3}  
    \resizebox{\linewidth}{!}{  
    \begin{tabular}{c|c|c|c|c|c|c|cccccccccccc}
    \hline
    \xrowht{10pt} 
    \multirow{2}{*}{ID} & \multirow{2}{*}{Model} & 
    \multirow{2}{*}{\shortstack{Method}} & 
    \multirow{2}{*}{\shortstack{Train\\set}} & 
    \multirow{2}{*}{\shortstack{Granularity}}  & 
    \#Params &GFLOPs$\downarrow$ & 
    ASR & KS & IC & PR & ER & QbE & SF & SID & ASV & SD & \\
    \cline{6-18}
    \xrowht{7pt}
    & & & & & Millions  & Total / Trans. & WER$\downarrow$ & Acc$\uparrow$ & Acc$\uparrow$ & PER$\downarrow$ & Acc$\uparrow$ & MTWV$\uparrow$ & F1$\uparrow$ / CER$\downarrow$ & Acc$\uparrow$ & EER$\downarrow$ & DER$\downarrow$ \\
    \hline
    1 & \multicolumn{4}{c|}{wav2vec2.0 Base} & 95.04 &14.01 / 8.59  &6.43 &96.23 &92.35 &5.74 &63.43 &0.0233 &88.30 / 24.77 &75.18 & 6.02 &6.08  \\ 
    \hline
    2 & \multirow{10}{*}{wav2vec2.0 Base} & UMP & \multirow{5}{*}{ls 100h} & \multirow{3}{*}{Channel} & 23.59 &6.86 / 1.44 & 14.50 & 95.16 & 92.20 & 14.52 & 61.61 & 0.0304 & 79.29 / 39.51 & 62.55 & 7.90 & 6.68\\
    3 & & DPHuBERT & &  & 23.59 &6.87 / 1.45 &13.29&95.00&93.30&11.04 &60.67&0.0427&83.46 / 34.18 & 67.88&7.44 &6.45\\
    4 & & {SASPG} & &  & 23.59 &6.86 / 1.44 & \cellcolor{cyan!20}12.91$^*$ &94.81&\cellcolor{cyan!20}93.38$^*$&\cellcolor{cyan!20}11.61 &\cellcolor{cyan!20}\textbf{62.05}$^*$&\cellcolor{cyan!20}\textbf{0.0495}$^*$&\cellcolor{cyan!20}81.69 / 35.62 &\cellcolor{cyan!20}69.49$^*$&\cellcolor{cyan!20}7.04$^*$& \cellcolor{cyan!20}6.65\\
    \cline{3-3}
    \cline{5-7}
    5 & & UMP &  & \multirow{2}{*}{Node} & 23.56 &\multirow{2}{*}{14.01 / 8.59} & 16.54 &94.55 &88.06 &18.65 &59.47&\underline{0.0329}&78.44 / 42.02 &59.75 &8.77&6.84\\
    6 &&SASPG & & & 23.56 & & \cellcolor{cyan!20}\textbf{\underline{8.22}}&\cellcolor{cyan!20}\textbf{\underline{95.85}}&\cellcolor{cyan!20}\textbf{\underline{95.81}}&\cellcolor{cyan!20}\textbf{\underline{7.63}} &\cellcolor{cyan!20}61.64&\cellcolor{cyan!20}0.0475& \cellcolor{cyan!20}\textbf{\underline{85.97} / \underline{28.24}} &\cellcolor{cyan!20}\textbf{\underline{75.15}} &\cellcolor{cyan!20}\textbf{\underline{6.11}}& \cellcolor{cyan!20}\underline{6.51}\\ 
    \cline{3-17}
    7 & & UMP & \multirow{5}{*}{ls 960h} & \multirow{3}{*}{Channel} & 23.63 &6.86 / 1.44 &13.61 & 95.81  &92.64&13.71&60.89&0.0405&83.98 / 33.10 &65.28&7.65&6.84\\
    8 & & DPHuBERT & &  & 23.63 &6.87 / 1.45  &  12.75 & 95.33&93.01&11.19 &60.32&0.0481&85.52 / 30.85&72.72&6.59&6.40\\
    9 & & SASPG & &  & 23.62  &6.86 / 1.44 &  \cellcolor{cyan!20}12.25$^*$ & 95.13&\cellcolor{cyan!20}93.12$^*$&\cellcolor{cyan!20}11.87 &\cellcolor{cyan!20}61.50$^*$& \cellcolor{cyan!20}0.0516$^*$&\cellcolor{cyan!20}85.23 / 30.46$^*$ &\cellcolor{cyan!20}74.00$^*$&\cellcolor{cyan!20}\textbf{6.26}$^*$& \cellcolor{cyan!20}6.66\\ 
    \cline{3-3}
    \cline{5-7}
    10 & & UMP & &\multirow{2}{*}{Node}  & 23.56  &\multirow{2}{*}{14.01 / 8.59} &15.42&95.16&89.64 &17.98 &\underline{61.30}&0.0383&80.63 / 37.93&62.32 &8.14&7.34\\ 
    11 &  & SASPG& & & 23.56 & & \cellcolor{cyan!20}\textbf{\underline{7.65}} &\cellcolor{cyan!20}\textbf{\underline{96.04}}&\cellcolor{cyan!20}\textbf{\underline{94.94}}&\cellcolor{cyan!20}\textbf{\underline{7.30}} &\cellcolor{cyan!20}\textbf{\underline{62.24}}&\cellcolor{cyan!20}\textbf{\underline{0.0519}}&\cellcolor{cyan!20}\textbf{\underline{87.26} / \underline{25.90}} & \cellcolor{cyan!20}\textbf{\underline{75.54}}&\cellcolor{cyan!20}6.50&\cellcolor{cyan!20}\underline{6.41}\\ 
    \hline
    12 &\multicolumn{4}{c|}{HuBERT Base} & 94.68 &14.01 / 8.59 & 6.42 & 96.30  & 98.34 & 5.41 & 64.92 & 0.0736 & 88.53 / 25.20 & 81.42 & 5.11 & 5.88 \\ 
    \hline
    13 & \multirow{10}{*}{HuBERT Base} & UMP & \multirow{5}{*}{ls 100h} & \multirow{3}{*}{Channel} & 23.58 &6.86 / 1.44 & 13.84 &95.88&95.33 &12.75 &60.16&0.0540&82.16 / 35.81&68.50&7.71 &6.56 \\ 
    14 & & DPHuBERT &  &  & 23.58 &6.86 / 1.44  & 11.54  & 95.85 & 97.50  & 10.08 & 62.22 & 0.0624  & 84.97 / 31.32 & 73.09  & 6.09  & 6.28 \\ 
    15 && {SASPG} & &  & 23.56 &6.86 / 1.44 & \cellcolor{cyan!20}11.46$^*$  & 95.62  & \cellcolor{cyan!20}97.86$^*$  & \cellcolor{cyan!20}10.19  & \cellcolor{cyan!20}62.79$^*$  & 0.0537  & \cellcolor{cyan!20}84.22 / 32.11 & \cellcolor{cyan!20}75.25$^*$  & \cellcolor{cyan!20}6.41  & \cellcolor{cyan!20}6.21$^*$  \\ 
    \cline{3-3}
    \cline{5-7}
    16 & & UMP & & \multirow{2}{*}{Node} & 23.56 &\multirow{2}{*}{14.01 / 8.59} &15.13&95.23&\underline{95.57}&14.12&\underline{61.57}&0.0486&81.49 / 36.80&66.25&\underline{7.09}&7.05\\ 
    17 & & SASPG & &  & 23.56 & & \cellcolor{cyan!20}\textbf{\underline{9.61}}  & \cellcolor{cyan!20}\textbf{\underline{96.62}}  & \cellcolor{cyan!20}\textbf{\underline{97.92}}  & \cellcolor{cyan!20}\textbf{\underline{8.67}}  & \cellcolor{cyan!20}\textbf{\underline{63.94}}  & \cellcolor{cyan!20}\underline{0.0575}  & \cellcolor{cyan!20}\textbf{\underline{86.45}} / \cellcolor{cyan!20}\textbf{\underline{29.52}} & \cellcolor{cyan!20}\textbf{\underline{79.37}}  & \cellcolor{cyan!20}\underline{6.16}  & \cellcolor{cyan!20}\textbf{\underline{5.97}} \\ 
    \cline{3-17}
    18 & & UMP & \multirow{5}{*}{ls 960h} & \multirow{3}{*}{Channel} & 23.58  &6.86 / 1.44 &12.58 &95.94 & 97.07 &12.29&63.54&0.0610&85.00 / 31.68 &73.48&6.84&6.61\\ 
    19 &  & DPHuBERT & & & 23.58 &6.87 / 1.44 & 10.93  & 96.33  & 97.23  & 9.94  & 62.85  & 0.0701  & 87.04 / 27.97 & 75.52  & 6.35  & 6.13  \\ 
    20 & & {SASPG} & &  & 23.57 &6.86 / 1.43 & \cellcolor{cyan!20}10.89$^*$  & \cellcolor{cyan!20}96.07  & \cellcolor{cyan!20}97.68$^*$  & \cellcolor{cyan!20}9.80$^*$  & 61.87  & \cellcolor{cyan!20}0.0653  & \cellcolor{cyan!20}86.17 / 27.89$^*$  & \cellcolor{cyan!20}77.18$^*$  & \cellcolor{cyan!20}6.27$^*$  & \cellcolor{cyan!20}6.28  \\ 
    \cline{3-3}
    \cline{5-7}
    21 & &  UMP & & \multirow{2}{*}{Node}  & 23.56 &\multirow{2}{*}{14.01 / 8.59} &13.51&\underline{96.07}&96.84&13.56&62.55 &0.0565&84.76 / 32.15 &68.51 &\underline{6.41}&6.95\\ 
    22 & & SASPG & &  & 23.56 & & \cellcolor{cyan!20}\textbf{\underline{9.20}}  & \cellcolor{cyan!20}\underline{96.27}  & \cellcolor{cyan!20}\textbf{\underline{98.02}}  & \cellcolor{cyan!20}\textbf{\underline{8.57}}  & \cellcolor{cyan!20}\textbf{\underline{63.79}}  & \cellcolor{cyan!20}0.0602  & \cellcolor{cyan!20}\textbf{\underline{87.05}} / \cellcolor{cyan!20}\textbf{\underline{26.77}} & \cellcolor{cyan!20}\textbf{\underline{79.20}}  & \cellcolor{cyan!20}\textbf{\underline{5.75}}  & \cellcolor{cyan!20}\textbf{\underline{5.97}}  \\ 
    \hline
    23 &\multicolumn{4}{c|}{WavLM Base+} & 94.70 &14.01 / 8.59 &5.59 & 97.37 &99.00 &3.92 &68.65 &0.0988 &90.58 / 21.20 &89.42 &4.07 &3.50 \\
    \hline
    24 &\multirow{10}{*}{WavLM Base+} & UMP & \multirow{5}{*}{ls 100h} & \multirow{3}{*}{Channel} & 23.58 &6.86 / 1.43  &13.30 &96.07&97.39&9.99&62.89&0.0573&84.14 / 32.67&70.80&6.51&6.11\\
    25 & & DPHuBERT & & & 23.58 &6.86 / 1.44 & 11.32  & 96.23  & 98.23  & 8.77  & 63.50  & 0.0669  & 85.34 / 30.91 & 79.47  & 5.59  & 5.62 \\
    26 & & {SASPG} & & & 23.57 &6.85 / 1.43 & \cellcolor{cyan!20}11.23$^*$ & \cellcolor{cyan!20}96.20  & \cellcolor{cyan!20}98.26$^*$ & \cellcolor{cyan!20}9.13  & \cellcolor{cyan!20}63.43  & \cellcolor{cyan!20}0.0688$^*$  & \cellcolor{cyan!20}85.51$^*$ / 30.93 & \cellcolor{cyan!20}77.26  & \cellcolor{cyan!20}5.97  & \cellcolor{cyan!20}5.79  \\
    \cline{3-3}
    \cline{5-7}
    27 & & UMP & & \multirow{2}{*}{Node} & 23.56 &\multirow{2}{*}{14.01 / 8.59} &14.27&95.29&\underline{97.84}&11.05&62.21&0.0540 &83.17 / 34.49&69.86&\underline{6.33} &6.36\\
    28 & & SASPG & & & 23.56 & & \cellcolor{cyan!20}\textbf{\underline{8.92}}  & \cellcolor{cyan!20}\textbf{\underline{97.21}}  & \cellcolor{cyan!20}\textbf{\underline{98.73}}  & \cellcolor{cyan!20}\textbf{\underline{6.79}}  & \cellcolor{cyan!20}\textbf{\underline{66.61}}  & \cellcolor{cyan!20}\textbf{\underline{0.0846}}  & \cellcolor{cyan!20}\textbf{\underline{87.04}} / \cellcolor{cyan!20}\textbf{\underline{26.86}} & \cellcolor{cyan!20}\textbf{\underline{85.21}}  & \cellcolor{cyan!20}\textbf{\underline{4.94}}  & \cellcolor{cyan!20}\textbf{\underline{5.06}}  \\
    \cline{3-17}
    29 & & UMP & \multirow{5}{*}{ls 960h} &  \multirow{3}{*}{Channel} & 23.59 &6.86 / 1.44  &12.26 &96.46 &97.97&9.58&63.51&0.0727&86.12 / 27.66&71.55&6.67&5.80\\
    30 & & DPHuBERT & & & 23.59 &6.86 / 1.44 & 10.22  & 96.69  & 98.68  & 8.44  & 64.55  & 0.0834  & 87.96 / 25.60 & 81.26  & 5.51  & 5.68  \\
    31 & & {SASPG} & &  & 23.58 &6.86 / 1.43 & \cellcolor{cyan!20}10.28  & \cellcolor{cyan!20}96.62  & \cellcolor{cyan!20}98.00  & \cellcolor{cyan!20}8.76  & \cellcolor{cyan!20}64.27  & \cellcolor{cyan!20}0.0839$^*$  &\cellcolor{cyan!20}\textbf{88.15}$^*$ / 25.94 & \cellcolor{cyan!20}80.40  & \cellcolor{cyan!20}5.62  & \cellcolor{cyan!20}5.58$^*$  \\
    \cline{3-3}
    \cline{5-7}
    32 & & UMP & & \multirow{2}{*}{Node} & 23.56 &\multirow{2}{*}{14.01 / 8.59}  &12.65 &\underline{96.72} &97.65&10.28 &62.33 &0.0676&\underline{87.51} / 27.80&\underline{71.65}&\underline{6.29}&6.11\\ 
    33 & &SASPG & & & 23.56 & & \cellcolor{cyan!20}\textbf{\underline{8.45}}  & \cellcolor{cyan!20}\textbf{\underline{97.01}}  & \cellcolor{cyan!20}\underline{98.52}  & \cellcolor{cyan!20}\textbf{\underline{6.65}}  & \cellcolor{cyan!20}\textbf{\underline{66.94}}  & \cellcolor{cyan!20}\textbf{\underline{0.0884}}  & \cellcolor{cyan!20}87.82 / \cellcolor{cyan!20}{\textbf{\underline{24.26}}} & \cellcolor{cyan!20}\textbf{\underline{85.66}}  & \cellcolor{cyan!20}\textbf{\underline{5.21}}  & \cellcolor{cyan!20}\textbf{\underline{4.89}}  \\ 
    \hline
    34 &\multicolumn{4}{c|}{HuBERT Large} & 316.60 &36.25 / 30.44 & 3.62 & 95.29 & 98.76 & 3.53 & 67.62 & 0.0353 & 89.81 / 21.76 & 90.33 & 5.98 & 5.75 \\
    \hline
    35 & \multirow{10}{*}{HuBERT Large} & UMP & \multirow{5}{*}{ls 100h} & \multirow{3}{*}{Channel}  & 93.67 &13.90 / 8.10 &11.20&96.79&98.02&7.46&63.62 &0.0547&85.80 / 31.40&74.96&6.95&6.64\\
    36 & & DPHuBERT &  &   & 93.67 &13.93 / 8.12 & 7.35  & 96.30  & 98.21  & 5.04  & 65.04  & 0.0622  & 86.32 / 28.85 & 77.60  & 6.55  & 5.95 \\
    37 & & {SASPG} & & & 93.66 &13.91 / 8.10 & \cellcolor{cyan!20}7.55  & 96.69$^*$  & \cellcolor{cyan!20}98.68$^*$  & \cellcolor{cyan!20}5.61  & \cellcolor{cyan!20}65.48$^*$  & \cellcolor{cyan!20}\textbf{0.0630}$^*$  & \cellcolor{cyan!20}86.42$^*$ / 28.11$^*$ & \cellcolor{cyan!20}79.19$^*$  & \cellcolor{cyan!20}6.71  & \cellcolor{cyan!20}6.12  \\
    \cline{3-3}
    \cline{5-7}
    38 & &UMP & &\multirow{2}{*}{Node} & 93.53 &\multirow{2}{*}{36.25 / 30.44} &13.29&95.52&97.81&8.66&61.59&0.0505&84.36 / 33.23&67.46&8.02 &7.03\\
    39 & &SASPG & & & 93.53 & & \cellcolor{cyan!20}\textbf{\underline{4.74}}  & \cellcolor{cyan!20}95.85 & \cellcolor{cyan!20}\textbf{\underline{98.71}}  & \cellcolor{cyan!20}\textbf{\underline{4.21}}  & \cellcolor{cyan!20}\textbf{\underline{66.64}}  & 0.0391 & \cellcolor{cyan!20}\textbf{\underline{87.83}} / \cellcolor{cyan!20}\textbf{\underline{25.38}} & \cellcolor{cyan!20}\textbf{\underline{85.90}} & \cellcolor{cyan!20}\textbf{\underline{6.26}}  & \cellcolor{cyan!20}\textbf{\underline{5.84}}  \\
    \cline{3-17}
    40 & & UMP & \multirow{5}{*}{ls 960h} & \multirow{3}{*}{Channel} & 94.57 & 13.99 / 8.19 &7.94&97.31&99.05&6.00&65.40&0.0613&88.04 / 25.58 &80.74 &7.16 &6.02\\
    41 &  & DPHuBERT &  &  & 94.57 & 14.02 / 8.22 & 6.22  & 94.84  & 98.52  & 4.69  & 65.55  & 0.0270  & 88.34 / 24.76 & 80.90  & 7.71  & 5.83  \\
    42 &  & {SASPG} & &   & 94.38 & 13.99 / 8.18 & \cellcolor{cyan!20}6.68  & 96.66$^*$  & 98.50  & \cellcolor{cyan!20}5.09  & 65.38  & \cellcolor{cyan!20}\textbf{0.0718}$^*$  & \cellcolor{cyan!20}88.80$^*$ / 24.35$^*$ & \cellcolor{cyan!20}82.67$^*$  & \cellcolor{cyan!20}6.83$^*$ & \cellcolor{cyan!20}5.95  \\
    \cline{3-3}
    \cline{5-7}
    43 &  & UMP & & \multirow{2}{*}{Node} & 94.37 &\multirow{2}{*}{36.25 / 30.44}  &9.11&97.11&98.68 &7.18&65.07&\underline{0.0656}&\underline{88.08} / 26.35&75.51&7.76&6.34 \\ 
    44 &  & SASPG & & & 94.37 & & \cellcolor{cyan!20}\textbf{\underline{4.25}}  & 95.81  & \underline{\cellcolor{cyan!20}98.87}  & \cellcolor{cyan!20}\textbf{\underline{4.09}}  & \cellcolor{cyan!20}\textbf{\underline{66.11}}  & 0.0374  & \cellcolor{cyan!20}\textbf{\underline{90.14}} / \cellcolor{cyan!20}\textbf{\underline{22.61}} & \cellcolor{cyan!20}\textbf{\underline{86.38}}  & \cellcolor{cyan!20}\textbf{\underline{6.42}}  & \cellcolor{cyan!20}\textbf{\underline{5.70}}  \\ 
    \hline
    \end{tabular} }

    \label{tab:superb}
\end{table*}
\subsection{Performance Analysis on the SUPERB Benchmark}

To overcome the restriction of single-task applications, we evaluate our method using task-agnostic pruning via distillation on the SUPERB benchmark (Table~\ref{tab:superb}). We organized our analysis around two core dimensions: fine-grained unstructured pruning and coarse-grained structured pruning.

\subsubsection{Performance-Oriented Node-level SASPG}

\textbf{i) Node-level SASPG vs. Node-level UMP.} Extending the advantage of dynamic consistency to task-agnostic distillation, node-level SASPG systematically outperforms the node-level offline UMP baseline (\textbf{blue-colored cells}).
\textbf{ii) Node-level SASPG vs. Channel-level SASPG.} Unlike node-level offline UMP, which suffers catastrophic representation collapse and falls behind its structured counterpart, our node-level SASPG predominantly surpasses the highly competitive channel-level SASPG (\textbf{underlined results}).
\textbf{iii) Node-level SASPG vs. other methods.} Node-level SASPG achieves \syncfix{the best performance among the compressed configurations in Table~\ref{tab:superb}} (\textbf{bold results}) across \textbf{7}/\textbf{8} (ID~6/ID~11), \textbf{8}/\textbf{8} (ID~17/ID~22), \textbf{10}/\textbf{8} (ID~28/ID~33), and \textbf{8}/\textbf{7} (ID~39/ID~44) out of \textbf{10} downstream tasks for wav2vec2.0-Base, HuBERT-Base, WavLM-Base+, and HuBERT-Large under the \textbf{100h}/\textbf{960h} settings, respectively. When combined with its channel-level configurations, SASPG's performance superiority expands to \textbf{9}/\textbf{9} (ID~4 and ID~6/ID~9 and ID~11), \textbf{8}/\textbf{8} (ID~15 and ID~17/ID~20 and ID~22), \textbf{10}/\textbf{9} (ID~26 and ID~28/ID~31 and ID~33), and \textbf{9}/\textbf{8} (ID~37 and ID~39/ID~42 and ID~44) tasks across the four respective models under the \textbf{100h}/\textbf{960h} settings.

\subsubsection{Acceleration-oriented Channel-level SASPG}

\textbf{i) Channel-level SASPG vs. Channel-level UMP.} Channel-level SASPG largely outperforms its channel-level offline UMP counterpart across the vast majority of evaluated tasks (\textbf{blue-colored cells}).
\textbf{ii) Channel-level SASPG vs. Channel-level DPHuBERT.} 
       In terms of pruning-specific parameters in Table~\ref{tab:param_feature}, it delivers extreme \textbf{1542$\times$} (\textbf{37008}/\textbf{24}) and \textbf{2056$\times$} (\textbf{98688}/\textbf{48}) reductions. 
       Maintaining this parameter efficiency, it surpasses coarse-grained channel-level DPHuBERT on specific subtasks (\textbf{Asterisked ($*$) results}). Specifically, it outperforms coarse-grained channel-level DPHuBERT across \textbf{6}/\textbf{6} (ID~4 vs. ID~3/ID~9 vs. ID~8), \textbf{5}/\textbf{5} (ID~15 vs. ID~14/ID~20 vs. ID~19), \textbf{3}/\textbf{2} (ID~26 vs. ID~25/ID~31 vs. ID~30), and \textbf{6}/\textbf{5} (ID~37 vs. ID~36/ID~42 vs. ID~41) out of \textbf{10} subtasks for wav2vec2.0-Base, HuBERT-Base, WavLM-Base+, and HuBERT-Large under the \textbf{100h}/\textbf{960h} settings, respectively.

\Needspace{4\baselineskip}\noindent\revision{Table~\ref{tab:multiseed} reports three-seed results for ASR-specific pruning at 70\% sparsity and task-agnostic pruning on five SUPERB tasks. Each task-agnostic repetition uses the same seed for both distillation stages and downstream training.}

\begin{table}[!t]
\caption{\revision{Representative SASPG results, reported as mean $\pm$ sample standard deviation (\%) over seeds 1000, 2022, and 3407. For emotion recognition (ER), each seed's score is averaged over five folds.}}
\label{tab:multiseed}
\centering
\fontsize{7}{8.4}\selectfont
\setlength{\tabcolsep}{3pt}
\renewcommand{\arraystretch}{1.1}
\begin{tabular*}{\columnwidth}{@{\extracolsep{\fill}}cccccc@{}}
\hline
\multicolumn{6}{c}{\revision{ASR-specific pruning: 70\% sparsity}} \\
\hline
\revision{ID} & \revision{\makecell{Train\\set}} & \revision{Granularity} & \revision{Penalty} & \revision{\shortstack{test-clean\\WER $\downarrow$}} & \revision{\shortstack{test-other\\WER $\downarrow$}} \\
\hline
\multicolumn{6}{c}{\revision{wav2vec2.0-Base}} \\
\hline
\revision{1} & \revision{100h} & \revision{Node} & \revision{DP} & \revision{\seedresult{6.41}{0.07}} & \revision{\seedresult{17.55}{0.29}} \\
\revision{2} & \revision{100h} & \revision{Node} & \revision{AL} & \revision{\seedresult{6.26}{0.07}} & \revision{\seedresult{17.13}{0.25}} \\
\revision{3} & \revision{100h} & \revision{Channel} & \revision{DP} & \revision{\seedresult{8.20}{0.27}} & \revision{\seedresult{23.59}{0.88}} \\
\revision{4} & \revision{100h} & \revision{Channel} & \revision{AL} & \revision{\seedresult{8.26}{0.16}} & \revision{\seedresult{23.23}{0.57}} \\
\hline
\revision{5} & \revision{960h} & \revision{Node} & \revision{DP} & \revision{\seedresult{4.93}{0.05}} & \revision{\seedresult{12.15}{0.22}} \\
\revision{6} & \revision{960h} & \revision{Node} & \revision{AL} & \revision{\seedresult{5.05}{0.07}} & \revision{\seedresult{12.18}{0.12}} \\
\revision{7} & \revision{960h} & \revision{Channel} & \revision{DP} & \revision{\seedresult{7.43}{0.23}} & \revision{\seedresult{17.37}{0.68}} \\
\revision{8} & \revision{960h} & \revision{Channel} & \revision{AL} & \revision{\seedresult{8.15}{0.29}} & \revision{\seedresult{18.21}{0.42}} \\
\hline
\multicolumn{6}{c}{\revision{HuBERT-Large}} \\
\hline
\revision{9} & \revision{100h} & \revision{Node} & \revision{DP} & \revision{\seedresult{3.70}{0.07}} & \revision{\seedresult{10.24}{0.12}} \\
\revision{10} & \revision{100h} & \revision{Node} & \revision{AL} & \revision{\seedresult{3.87}{0.05}} & \revision{\seedresult{10.21}{0.06}} \\
\revision{11} & \revision{100h} & \revision{Channel} & \revision{DP} & \revision{\seedresult{6.10}{0.47}} & \revision{\seedresult{19.97}{1.52}} \\
\revision{12} & \revision{100h} & \revision{Channel} & \revision{AL} & \revision{\seedresult{6.04}{0.06}} & \revision{\seedresult{19.90}{0.30}} \\
\hline
\revision{13} & \revision{960h} & \revision{Node} & \revision{DP} & \revision{\seedresult{3.01}{0.09}} & \revision{\seedresult{6.94}{0.14}} \\
\revision{14} & \revision{960h} & \revision{Node} & \revision{AL} & \revision{\seedresult{3.09}{0.01}} & \revision{\seedresult{7.24}{0.06}} \\
\revision{15} & \revision{960h} & \revision{Channel} & \revision{DP} & \revision{\seedresult{5.32}{0.15}} & \revision{\seedresult{13.44}{0.44}} \\
\revision{16} & \revision{960h} & \revision{Channel} & \revision{AL} & \revision{\seedresult{5.73}{0.15}} & \revision{\seedresult{14.28}{0.29}} \\
\hline
\end{tabular*}
\par\vspace{2pt}
\begin{tabular*}{\columnwidth}{@{\extracolsep{\fill}}cccc@{}}
\hline
\multicolumn{4}{c}{\revision{\shortstack{Task-agnostic pruning: LibriSpeech 100h\\Channel-level, \revision{75\% model-parameter reduction}}}} \\
\hline
\revision{ID} & \revision{Metric} & \revision{HuBERT-Base} & \revision{WavLM-Base+} \\
\hline
\revision{17} & \revision{ASR WER (test-clean) $\downarrow$} & \revision{\seedresult{11.41}{0.18}} & \revision{\seedresult{11.18}{0.18}} \\
\revision{18} & \revision{KS accuracy $\uparrow$} & \revision{\seedresult{95.65}{0.07}} & \revision{\seedresult{96.25}{0.07}} \\
\revision{19} & \revision{IC accuracy $\uparrow$} & \revision{\seedresult{97.88}{0.05}} & \revision{\seedresult{98.28}{0.05}} \\
\revision{20} & \revision{PR PER $\downarrow$} & \revision{\seedresult{10.22}{0.04}} & \revision{\seedresult{9.14}{0.04}} \\
\revision{21} & \revision{ER accuracy $\uparrow$} & \revision{\seedresult{62.85}{0.18}} & \revision{\seedresult{63.49}{0.18}} \\
\hline
\end{tabular*}
\end{table}

\begin{table}[htbp]
\caption{ASR performance comparison with previous unstructured pruning methods on the LibriSpeech test-clean subset. Values in parentheses denote sparsity. Bold WER values indicate lower WER than the corresponding previous pruning method for the same model and sparsity.}
    \centering
    \setlength\tabcolsep{1pt}
    \setlength{\extrarowheight}{2pt} 
    \resizebox{\linewidth}{!}{
    \begin{tabular}{c|c|c|c|c|c|c|c}
        \hline
        \multirow{2}{*}{ID} & \multirow{2}{*}{System} & \multirow{2}{*}{\shortstack{Train\\set}} & \multirow{2}{*}{\shortstack{\# Params\\(M)}} & \multicolumn{3}{c|}{Training time \& config} & \multirow{2}{*}{WER\%}\\
        \cline{5-7}
         & & & & Hours& Epochs& Iterations&\\
         \hline
         1 & wav2vec2.0-Base & & 95.04 &\multicolumn{3}{c|}{-} &6.06 \\
         \hline
        2 & SparseWAV2VEC2 (85\%)~\cite{gu2024sparsewav} &\multirow{4}{*}{100h} &22.27  &- &\multirow{2}{*}{$>$40} &- & 8.08\\
        3 & SparseWAV2VEC2 (75\%)~\cite{gu2024sparsewav} & &-  &- & &-  &7.11 \\
        \cline{1-2} \cline{4-8}
        4 & SASPG wav2vec2.0-Base (85\%) &  &22.20 & \multirow{2}{*}{11} &\multirow{2}{*}{30} &\multirow{2}{*}{27K}&\textbf{7.46}\\
         5 & SASPG wav2vec2.0-Base (75\%) & &30.70 & & & &\textbf{6.71} \\
        \hline
        6 & WavLM-Base+ & &94.70 &\multicolumn{3}{c|}{-} &\syncfix{5.59} \\
        \hline
         7 & SparseWAVLM+ (85\%)~\cite{gu2024sparsewav} & \multirow{2}{*}{100h}&22.98 &- &{$>$40} &- & 7.12\\
        \cline{1-2} \cline{4-8}
        8 & SASPG WavLM-Base+ (85\%) && 22.21 &11 &30 &27K &\textbf{7.05}\\
         \hline
    \end{tabular}      
    }
    \vspace{-0.4cm}
    \label{tab:sota unstr}

\end{table}

\begin{table*}[htbp]
\caption{Performance comparison with previous compression methods on the SUPERB benchmark. \textbf{Bold} and \underline{underlined} results indicate that our Channel-level SASPG achieves the best and second-best performance, respectively, for a given model and dataset setting (100h or 960h). \textcolor{black}{Tied scores share a rank; second-best denotes the next distinct score.} \textcolor{black}{Blue-colored cells}: Node-level SASPG strictly outperforms all evaluated channel-level methods.}
    \centering
    \setlength\tabcolsep{3pt}  
    \renewcommand{\arraystretch}{1.3}  
    \resizebox{\linewidth}{!}{  
    \begin{tabular}{c|c|c|c|c|c|ccccccccccccc}
    \hline
    \xrowht{10pt} 
    \multirow{2}{*}{ID} & \multirow{2}{*}{Model} & 
    \multirow{2}{*}{\shortstack{Method}} & 
    \multirow{2}{*}{\shortstack{Train\\set}} & 
    \multirow{2}{*}{\shortstack{Granularity}}  & 
    \#Params & 
    ASR & KS & IC & PR & ER & QbE & SF & SID & ASV & SD & \\
    \cline{6-17}
    \xrowht{7pt}
    & & & & & Millions & WER$\downarrow$ & Acc$\uparrow$ & Acc$\uparrow$ & PER$\downarrow$ & Acc$\uparrow$ & MTWV$\uparrow$ & F1$\uparrow$ / CER$\downarrow$ & Acc$\uparrow$ & EER$\downarrow$ & DER$\downarrow$ \\
    \hline
    1 & \multicolumn{4}{c|}{wav2vec2.0 Base} & 95.04  &6.43 &96.23 &92.35 &5.74 &63.43 &0.0233 &88.30 / 24.77 &75.18 & 6.02 &6.08  \\ 
    \hline
    2 & \multirow{6}{*}{wav2vec2.0 Base} & FitW2V2~\cite{fithubert} & \multirow{3}{*}{ls 100h} & \multirow{2}{*}{Channel} & 22.49 &14.77 &94.68 &90.03 &16.50 &62.87 &0.0380 &81.95 / 34.74& 51.65 &7.43 &6.94\\
    \cline{3-3}
    3 && \multirow{2}{*}{SASPG} & &  & 23.59 & \textbf{12.91}  &\textbf{94.81}&\textbf{93.38}&\textbf{11.61} &\underline{62.05}&\textbf{0.0495}&\underline{81.69} / \underline{35.62} &\textbf{69.49} &\textbf{7.04}& \textbf{6.65}\\
    \cline{5-5}
    4 && & & Node & 23.56 & \cellcolor{cyan!20}8.22&\cellcolor{cyan!20}95.85&\cellcolor{cyan!20}95.81&\cellcolor{cyan!20}7.63 &61.64&0.0475& \cellcolor{cyan!20}85.97 / 28.24 &\cellcolor{cyan!20}75.15 &\cellcolor{cyan!20}6.11& \cellcolor{cyan!20}6.51\\ 
    \cline{3-17}
    5 & & FitW2V2~\cite{fithubert} & \multirow{3}{*}{ls 960h} & \multirow{2}{*}{Channel} & 31.63 & 11.44&96.04 &93.38 &12.22 &62.35 &0.0475 &86.65 / 29.40& 64.71& 6.65& 6.44\\
    \cline{3-3}
    6 & & \multirow{2}{*}{SASPG} & &  & 23.62 &  \underline{12.25} & \underline{95.13}&\underline{93.12}&\textbf{11.87} &\underline{61.50}& \textbf{0.0516}&\underline{85.23} / \underline{30.46} &\textbf{74.00}&\textbf{6.26}& \underline{6.66}\\ 
    \cline{5-5}
    7 &  & & & Node & 23.56 & \cellcolor{cyan!20}7.65 &96.04&\cellcolor{cyan!20}94.94&\cellcolor{cyan!20}7.30 &62.24&\cellcolor{cyan!20}0.0519&\cellcolor{cyan!20}87.26 / 25.90 & \cellcolor{cyan!20}75.54&6.50&\cellcolor{cyan!20}6.41\\ 
    \hline
    8 &\multicolumn{4}{c|}{HuBERT Base} & 94.68 & 6.42 & 96.30  & 98.34 & 5.41 & 64.92 & 0.0736 & 88.53 / 25.20 & 81.42 & 5.11 & 5.88 \\ 
    \hline
    9 & \multirow{16}{*}{HuBERT Base} & DPHuBERT~\cite{peng2023dphubert} & \multirow{5}{*}{ls 100h} & \multirow{4}{*}{Channel} & 23.57 & 11.38 & 96.36 & 97.42 & 10.02 & 62.78 & 0.0634 & 84.83 / 33.03 & 73.37 & 6.25 & 6.03 \\
    10 & &FitHuBERT~\cite{fithubert}  & &  & 22.49 &12.66 &96.23 &94.20 &14.05 & 61.67& 0.0579& 83.41 / 34.00 &54.24& 7.88& 7.19  \\ 
    11  & & ARMHuBERT-S\cite{jang2023recycleanddistill} & & &22.39 &11.83 &96.01 &95.97 &9.17  &63.29 &0.0569 &83.89 / 33.29 &66.48 &5.92 &6.23   \\
    12  & & STaRHuBERT~\cite{jang2024star} & & & 22.31 &10.92 & 96.46 &97.21 & 9.17 &61.22 &0.0626 &84.89 / 30.77 &72.26 &5.66 &5.97 \\
    \cline{3-3}
    13 && \multirow{2}{*}{SASPG} & &  & 23.56  & 11.46  & 95.62  & \textbf{97.86}  & 10.19  & \underline{62.79}  & 0.0537  & 84.22 / \underline{32.11} & \textbf{75.25}  & 6.41  & 6.21  \\ 
    \cline{5-5}
    14 & & & & Node & 23.56 & \cellcolor{cyan!20}9.61  & \cellcolor{cyan!20}96.62  & \cellcolor{cyan!20}97.92  & \cellcolor{cyan!20}8.67  & \cellcolor{cyan!20}63.94  & 0.0575  & \cellcolor{cyan!20}86.45 / 29.52 & \cellcolor{cyan!20}79.37  & 6.16  & 5.97 \\ 
    \cline{3-17}
    15 & & DistillHuBERT~\cite{peng2023dphubert} & \multirow{9}{*}{ls 960h} & \multirow{8}{*}{Channel} & 23.49 &13.37 &95.98& 94.99 &16.27 &63.02 &0.0511 &82.57 / 35.59 &73.54 &8.55 &6.19\\ 
    16 & & DPHuBERT~\cite{peng2023dphubert} &  & & 23.59 & 10.47 & 96.36 & 97.92 & 9.67 & 63.16 & 0.0693 & 86.86 / 28.26 & 76.83 & 5.84 & 5.92   \\ 
    17 & &FitHuBERT~\cite{fithubert}  & &  & 22.49 &12.09 & 96.27 &91.25 &13.32 &59.82 &0.0489 &84.06 / 32.46& 55.71& 8.00 &6.84 \\ 
    18 & &SKHuBERT~\cite{zampierin2024skillsimilarityawareknowledgedistillation}  & &  & 23.59 &10.78 &96.23 &97.53 &8.29 & 63.83& 0.0646 &86.68 / 27.45 &74.43 &6.14 &6.46\\
    19  & & DKD LSTM HuBERT~\cite{de2023lstmhubert} & & &18.80 &10.64 &96.95 &96.94 &8.57 &61.92 & 0.0535 & 85.81 / 30.73 & 56.71 &7.80 &6.48  \\ 
    20  & & 12-Layer Half~\cite{deepvswide} & & &26.87 &10.96 &97.24 &96.97 &10.67 & 63.24& 0.0604 &86.11 / 30.93& 69.52& 6.13& 6.81\\
    21  & & ARMHuBERT-S~\cite{jang2023recycleanddistill} & & &22.39 &10.82 &96.82 &97.02 &8.63 &62.96 & 0.0720 &86.34 / 29.02 &63.76 &5.58 &7.01  \\
    22  & & STaRHuBERT~\cite{jang2024star} & & & 22.31 &9.35 &96.27 &97.55 &8.16 &63.01 & 0.0688 & 87.94 / 25.31 & 77.58 & 5.39 & 6.05 \\ 
    \cline{3-3}
    23 & & \multirow{2}{*}{SASPG} & &  & 23.57 & 10.89  & 96.07  & \underline{97.68}  & 9.80  & 61.87  & 0.0653  & 86.17 / 27.89  & \underline{77.18}  & 6.27  & 6.28  \\ 
    \cline{5-5}
    24 & & & & Node  & 23.56 & \cellcolor{cyan!20}9.20  & 96.27  & \cellcolor{cyan!20}98.02  & 8.57  & 63.79  & 0.0602  & 87.05 / 26.77 & \cellcolor{cyan!20}79.20  & 5.75  & 5.97  \\ 
    \hline
    25 &\multicolumn{4}{c|}{WavLM Base+} & 94.70 &5.59 & 97.37 &99.00 &3.92 &68.65 &0.0988 &90.58 / 21.20 &89.42 &4.07 &3.50 \\
    \hline
    26 &\multirow{9}{*}{WavLM Base+} & ARMwavLM-S\cite{jang2023recycleanddistill} & \multirow{3}{*}{ls 100h} & \multirow{2}{*}{Channel} &22.39 &11.37  &96.30 &96.76 &8.33 &63.41 &0.0579 &84.89 / 31.95 &65.40 & 6.38 &7.41\\
    \cline{3-3}
    27 & & \multirow{2}{*}{SASPG} & & & 23.57  & \textbf{11.23} & \underline{96.20}  & \textbf{98.26} & \underline{9.13}  & \textbf{63.43}  & \textbf{0.0688}  & \textbf{85.51} / \textbf{30.93} & \textbf{77.26}  & \textbf{5.97}  & \textbf{5.79}  \\
    \cline{5-5}
    28 & & & & Node & 23.56 & \cellcolor{cyan!20}8.92  & \cellcolor{cyan!20}97.21  & \cellcolor{cyan!20}98.73  & \cellcolor{cyan!20}6.79  & \cellcolor{cyan!20}66.61  & \cellcolor{cyan!20}0.0846  & \cellcolor{cyan!20}87.04 / 26.86 & \cellcolor{cyan!20}85.21  & \cellcolor{cyan!20}4.94  & \cellcolor{cyan!20}5.06  \\
    \cline{3-17}
    29 & & DPWavLM~\cite{peng2023dphubert} & \multirow{6}{*}{ls 960h} &  \multirow{5}{*}{Channel} & 23.59 & 10.19 & 96.27 & 98.58 & 8.22 & 65.24 & 0.0874 & 87.68 / 26.11 & 82.11 & 5.98 & 5.53 \\
    30 & &SKWavLM~\cite{zampierin2024skillsimilarityawareknowledgedistillation}  & &  & 23.51 &10.03 & 96.75 &98.37 &6.90 &66.22& 0.0727 &88.62 / 25.36 &74.67& 5.67 &5.53\\
    31 & &Wang et al.~\cite{wang2023task}  & &  & 26.57  &10.61 &96.57 &98.08 &9.09 & 63.61 &0.0740 &87.14 / 27.13& 74.56& 6.17& 6.11\\
    32  & & ARMwavLM-S\cite{jang2023recycleanddistill} & & &22.39 &10.03 &97.01  &97.76 &7.42  &64.54 &0.0741 &87.41 / 26.97 &71.29 &5.99 &7.11  \\
    \cline{3-3}
    33 & & \multirow{2}{*}{SASPG} & &  & 23.58  & 10.28  & 96.62  & 98.00  & 8.76  & 64.27  & \underline{0.0839}  &\underline{88.15} / \underline{25.94} & \underline{80.40}  & \textbf{5.62}  & \textcolor{black}{\underline{5.58}}  \\
    \cline{5-5}
    34 & & & & Node & 23.56 & \cellcolor{cyan!20}8.45  & 97.01  & 98.52  & \cellcolor{cyan!20}6.65  & \cellcolor{cyan!20}66.94  & \cellcolor{cyan!20}0.0884  & 87.82 / \colorbox{cyan!20}{24.26} & \cellcolor{cyan!20}85.66  & \cellcolor{cyan!20}5.21  & \cellcolor{cyan!20}4.89  \\
    \hline
    35 &\multicolumn{4}{c|}{HuBERT Large} & 316.60 & 3.62 & 95.29 & 98.76 & 3.53 & 67.62 & 0.0353 & 89.81 / 21.76 & 90.33 & 5.98 & 5.75 \\
    \hline
    36 & \multirow{3}{*}{HuBERT Large} & DPHuBERT~\cite{peng2023dphubert} & \multirow{3}{*}{ls 960h} & \multirow{2}{*}{Channel} & 94.59 & 6.23 & 94.51 & 98.47 & 4.46 & 65.11 & 0.0246 & 88.37 / 24.60 & 83.17 & 7.05 & 5.79 \\
    \cline{3-3}
    37 & & \multirow{2}{*}{SASPG} & & & 94.38 & \underline{6.68} & \textbf{96.66} & \textbf{98.50} & \underline{5.09} & \textbf{65.38} & \textbf{0.0718} & \textbf{88.80} / \textbf{24.35} & \underline{82.67} & \textbf{6.83} & \underline{5.95} \\
    \cline{5-5}
    38 & & & & Node & 94.37 & \cellcolor{cyan!20}4.25 & 95.81 & \cellcolor{cyan!20}98.87 & \cellcolor{cyan!20}4.09 & \cellcolor{cyan!20}66.11 & 0.0374 & \cellcolor{cyan!20}90.14 / \cellcolor{cyan!20}22.61 & \cellcolor{cyan!20}86.38 & \cellcolor{cyan!20}6.42 & \cellcolor{cyan!20}5.70 \\
    \hline

    \end{tabular} }
    \label{tab:sota str}
\end{table*}

\subsection{Comparison with other compression methods}

\subsubsection{ASR-specific Unstructured Pruning}
As detailed in Table~\ref{tab:sota unstr}, \syncfix{we} report the WER alongside training efficiency metrics. Compared to existing unstructured pruning methods, node-level SASPG consistently achieves lower WERs at identical sparsity while requiring significantly fewer training epochs. This dual advantage in accuracy and training efficiency is robustly validated across multiple architectures, including wav2vec2.0-Base (ID~4-ID~5) and WavLM-Base+ (ID~8).

\subsubsection{Performance on the SUPERB Benchmark}
In {Table~\ref{tab:sota str}}, we define the ``best'' (\textbf{bold}) and ``second-best'' (\underline{underlined}) structured performance strictly among channel-level methods. {Blue-colored} cells highlight where our node-level SASPG strictly outperforms all evaluated channel-level baselines (ID~4, ID~7, ID~14, ID~24, ID~28, ID~34 and ID~38). \syncfix{Node-level SASPG is the unstructured method included in this task-agnostic comparison.}

For wav2vec2.0-Base, channel-level SASPG captures the best structured performance on \textbf{8} (ID~3, 100h) and \textbf{4} (ID~6, 960h) tasks, while the node-level variant secures the overall best performance on \textbf{8} (ID~4, 100h) and \textbf{7} (ID~7, 960h) tasks. On HuBERT-Base, despite exceptionally fierce competition, channel-level SASPG achieves the best or second-best structured results on \textbf{3} (ID~13, 100h) and \textbf{2} (ID~23, 960h) tasks, while the node-level model claims the best performance across \textbf{7} (ID~14, 100h) and \textbf{3} (ID~24, 960h) tasks. This superiority is further amplified on WavLM-Base+, where channel-level SASPG secures the best structured results on \textbf{8} (ID~27, 100h) tasks and the best or second-best results on \textcolor{black}{\textbf{5}} (ID~33, 960h) tasks, while node-level SASPG achieves \hnxu{the overall best performance on} \textbf{10} (ID~28, 100h) and \textbf{7} (ID~34, 960h) tasks. Finally, on the scaled-up HuBERT-Large, channel-level SASPG outperforms the original DPHuBERT across \textbf{6} (ID~37, 960h) tasks, and node-level SASPG outperforms all evaluated channel-level methods on \textbf{8} (ID~38, 960h) downstream tasks.

\section{Conclusion}
\label{sec7}

In this paper, we proposed Sparsity-Aware Self-Pinching Gates (SASPG), a high-performance and scalable framework for speech foundation model compression that tightly integrates model pruning and parameter update into a consistent cycle. By employing highly compact, layer-tied gates governed by a single learnable threshold, SASPG can be jointly optimized using either a task-specific CTC loss or a task-agnostic distillation objective. Our unified design seamlessly supports both fine-grained node-level unstructured pruning and coarse-grained channel-level structured pruning, providing flexible trade-offs between performance retention and hardware-friendly acceleration.

Extensive evaluations demonstrate SASPG's superiority over offline magnitude and coarse-grained baselines (e.g., NASP, DPHuBERT). For ASR, node-level SASPG achieves statistically lossless 5$\times$ and 3.3$\times$ \syncfix{compression} on wav2vec2.0-Base and HuBERT-Large, while its channel-level variant \revision{achieves measured speedups of up to 1.62$\times$ and 1.87$\times$, respectively}. On the SUPERB benchmark, SASPG outperforms existing baselines across the vast majority of tasks for four distinct SSL architectures. Crucially, SASPG secures these \syncfix{performance gains} with a 2056$\times$ reduction in pruning-specific parameter overhead versus DPHuBERT, establishing a highly streamlined and versatile compression paradigm.


\end{document}